# Primary Defect Production in Doped Iron Grain Boundaries during Low Energy Collision Cascades

Yang Zhang[1], Blas P. Uberuaga[2], Enrique Martinez Saez[3] and Jason R. Trelewicz[1,4,*]

[1]Department of Materials Science and Chemical Engineering, Stony Brook University, Stony Brook, NY 11794

[2]Materials Science and Technology Division, Los Alamos National Laboratory, Los Alamos, NM 87545

[3]Department of Materials Science and Engineering, Clemson University, Clemson, SC 29634

[4]Institute for Advanced Computational Science, Stony Brook University, Stony Brook, NY 11794

*Corresponding Author at Stony Brook University: Jason.trelewicz@stonybrook.edu

## Abstract

This study explores the intricate interactions between grain boundaries (GBs) and irradiation-induced defects in nanocrystalline iron, highlighting the role of dopants like copper. Utilizing molecular dynamics simulations, the research delineates how GB properties, such as GB energy and defect formation energies, influence the formation and evolution of defects in low energy collision cascades. It reveals that GBs not only augment defect production but also show a marked preference for interstitials over vacancies, a behavior significantly modulated by the cascade's proximity to the GB. The presence of dopants is shown to alter GB properties, affecting both the rate and type of defect production, thereby underscoring the complex interplay between GB characteristics, dopant elements, and defect dynamics. Moreover, the investigation uncovers that the structural characteristics of GBs play a crucial role in cascade evolution and defect generation, with certain GB configurations undergoing reconfiguration in response to cascades. For instance, the reconfiguration of one pure Fe twist GB suggests that GB geometry can significantly influence defect generation mechanisms. These findings point to the potential of GB engineering in developing materials with enhanced radiation tolerance, advocating for a nuanced approach to material design. By tailoring GB properties and selectively introducing dopant elements, materials can be optimized to exhibit superior resistance to radiation-induced damage, offering insights for applications in nuclear reactors and other radiation-prone environments.

## 1. Introduction

The development of radiation-resistant materials is crucial for the advancement of fission and fusion reactors [1-4]. These reactors subject the structural components to extreme and complex radiation conditions, which can cause displacement damages due to high neutron irradiation. As a result, various types of defects in the atomic structure are produced. To mitigate the impact of radiation damage and enhance irradiation tolerance, several methods have been proposed. These methods aim to minimize primary defect production, control defect diffusion to sinks, and promote defect annihilation [5-8]. Several investigations have indicated that nanocrystalline (NC) metals with a high density of grain boundaries (GBs) may demonstrate improved radiation tolerance [9-14], as GBs can serve as sinks for defects and enhance defect recombination efficiency [15-18]. In contrast, comparisons of NC Au and polycrystalline (PC) Au from some reports [19, 20] have shown that NC Au may be more susceptible to radiation damage in terms of resistivity than the conventional polycrystal material at ~15K, whereas lower resistivity changes were observed in NC Au than in PC Au after irradiation at 300K. Resistivity increase signifies an increase in defect density, which suggests from Ref. [19, 20] that the grain boundary plays a complex role in both the defect production and defect recovery stages. These reports enlighten the search of the radiation tolerant materials by uncovering the radiation defect production and evolution mechanisms near the grain boundary.

Prior investigations have revealed that the thermodynamics and kinetics governing defect behaviors within the interaction region of GBs, such as defect formation energy and defect migration energy, exhibit variations among different GB configurations [18, 21-23]. Computational simulations [2, 6, 21, 22, 24-30] have provided evidence that when a collision cascade occurs in close proximity to a GB, atomic displacements result in the generation of vacancies and interstitials in both the grain boundary region and the grain matrix region. In comparison to cascades taking place solely within the grain matrix without GBs, the overall count of primary Frenkel pairs (surviving Frenkel pairs prior to thermal recombination or recovery) may increase due to the interaction between the cascade and the GB. However, the number of interstitials in the grain matrix diminishes, signifying that the GB exhibits a selective or biased capturing effect on interstitials over vacancies when the collision cascade develops nearby. Furthermore, the inclusion of GBs has been found to influence the defect annihilation process following a cascade, as indicated by accelerated molecular dynamics (aMD) simulations [22]: defect annihilation can occur through a GB interstitial emission mechanism, in conjunction with the GB capturing vacancies during annealing.

The aforementioned study[18] highlights the influence of different GB structures on nearby defect behavior, supported by direct observations [31, 32] and other simulations [2, 6, 10, 21, 26]. And all of which demonstrate variations in defect behavior as the microscopic structure of the GB changes. The explanation for these variations can be attributed to the modifications in both GB energy [33-38] and defect migration energy, as well as defect formation energy within the GB [18, 35, 39], which are influenced by the GB's microstructure. These GB properties play a role in irradiation-GB interactions, affecting defect production and recombination processes. However, further research is required to establish a quantitative correlation between GB properties and the evolution of radiation damage.

While nanocrystalline materials with a significant fraction of GBs have shown potential for improved radiation tolerance, concerns arise regarding the structural stability of the non-equilibrium highly strained structure [40-47]. Nevertheless, research indicates that the addition of solute elements can impede the grain growth of nanocrystalline materials attributed to various mechanisms[42, 45, 48-56], including the increase in activation energy, pinning of grain boundaries by solute precipitates, reduction of the driving force determined by the grain boundary energy, and optimization of the microstructure of nanocrystalline materials. Consequently, the engineering of practical nanocrystalline materials for extreme conditions may involve the incorporation of suitable solute elements. One potential solute element for Fe-based nanocrystalline alloys is Cu [46, 57, 58]. Cu may form nano-sized precipitates in Fe grain boundaries through segregation, thereby reducing the grain boundary energy via the mixing rule [59, 60]. The influence of solute elements on the grain boundary is complicated, necessitating an assessment of the radiation tolerance of alloyed grain boundaries prior to practical applications.

To advance the exploration of radiation-tolerant material design, this study aims to investigate the influence of GB properties and dopant elements on radiation defect production and evolution. Molecular dynamics simulations were conducted to examine collision cascades near different GB configurations, both with and without dopant elements. Through systematic analysis, various mechanisms affecting defect evolution were elucidated. The impact of both energetics and structural characteristics on cascade defect evolution was studied, utilizing statistical data from a large number of cascade simulations. However, it is important to note that the conclusions drawn in this study are preliminary, as they are limited by time and computational resources. Further research is necessary to thoroughly test, refine, and complement the findings presented in this work.

## 2. Methods

To address the research objectives, a set of collision cascade simulations is conducted in bi-crystal structures featuring diverse grain boundary configurations. The simulation setup, depicted in Figure 1, involves initiating a collision cascade by accelerating an atom in the upper grain (Grain A) towards the grain boundary in a direction perpendicular to the boundary plane.

Following the initiation, the cascade evolves freely with mass and energy conservation until the energy becomes uniformly distributed throughout the structure. At the conclusion of the free evolution period, data on defect production is gathered.

The molecular dynamics simulations are conducted using the Large-scale Atomic/Molecular Massively Parallel Simulator (LAMMPS) [61], utilizing a suitably tuned interatomic interaction potential. The primary component of the potential employed in this study is an embedded-atom method (EAM) potential described in Ref. [62], which specifically focuses on interface and defect properties in Fe-Cu alloy. However, considering the high-energy conditions involved in the collision cascade simulations, modifications are made to the short-range interaction to accurately capture high-energy collisions between atoms. To account for the effects of stopping power on energy transfer cross-sections, the Ziegler-Biersack-Littmark correction (ZBL correction [63]) is implemented. Careful attention is given to the choice of cutoff radii for the ZBL correction to ensure that it does not influence the thermodynamic energetics of the material, such as defect formation energy, defect migration energy, and interface formation energy. In the BCC iron lattice, the distance between two atoms is 2.52 Å, and in interstitial configurations, the closest distance exceeds 2 Å. The cutoff radius for the ZBL repulsion is set to 0.4 Å, with a smooth transition from 0.4 Å to 0.7 Å. Beyond a distance of 0.7 Å between two atoms, the interatomic interaction is solely determined by the EAM potential described in Ref. [62]. This combination of the EAM potential and the ZBL correction allows for more reliable predictions of defect evolution during collision cascades.

### 2.1. Preparation of the simulation samples

In this study, the focus was on investigating the relationship between GB properties and radiation resistance by examining a selective collection of pure Fe GB configurations and their Cu-doped counterparts. The GB configurations studied were determined based on the Coincidence Site Lattice (CSL) theory, which categorizes them as Σ5<100>{130}, Σ9<110>{114}, Σ11<110>{113}, Σ13<100>{150} symmetric tilt GBs, and Σ5{100}, Σ9{110}, Σ11{110}, Σ13{100} twist GBs. These GBs were created using the misorientation angle and tilt or twist axis provided by the CSL theory to determine the GB planes. However, since the GB possesses five degrees of freedom, different microstates can arise due to displacements in the GB plane. To identify the minimal energy configuration, the γ-surface of each GB configuration was calculated following established methods [64, 65]. The pristine structure of each GB was determined by selecting the configuration with the lowest energy, as shown in Figure 2. The microstructure of the grain boundary exhibits significant variations among different configurations, characterized by distinct atomic volume distributions, which may influence their radiation resistance. Furthermore, based on the atomic volume distribution, the thickness of the GB can be estimated. Given a 2% atomic volume difference in the grain matrix (ranging from 11.52 Å$^3$ to 11.99 Å$^3$), the grain boundary thickness is estimated to be approximately 1 nm.

The preparation of the pure Fe structures for the collision cascade tests was finalized following a 1ns incubation period at 10K to alleviate artificially induced internal stress. These structures were employed to investigate the influence of GB structure on radiation resistance. Furthermore, the study explored the impact of dopants on radiation resistance. The doping process was realized through a combined Monte Carlo molecular dynamics (MCMD) routine, employing the variance constrained semi-grand-canonical (VC-SGC) MC function as proposed in Ref. [66]. In the present study, each MC swap occurred after 1000 MD steps, encompassing a total of 1000000 MD steps and 1000 MC swaps. During the MC swap, the function sought to alter the particle type between Fe and Cu. The magnitude of Cu doping within the structure was determined by the chemical potential (Δμ) setting in the MC algorithms. In the referenced work [66], the acceptance probability of substituting particles is determined by the potential energy difference (ΔU) before and after the substitution, in addition to the chemical potential (Δμ) value, expressed as (ΔU + Δμ). If (ΔU + Δμ) is negative, the acceptance probability is 1. Otherwise, the acceptance probability diminishes exponentially as the value of (ΔU + Δμ) increases according to the Boltzmann factor. In our study, the saturation point of GB doping was investigated by controlling the Cu composition through adjustments in the Δμ value, ranging from -0.56 eV to -0.80 eV with increments of -0.02 eV. Figure 3 illustrates the Σ5 twist structure at the conclusion of each MCMD process, showcasing the trends in Cu composition within the GB region and the GB energy. Supplementary Figure 1 presents the corresponding trends for other GB configurations. With a relatively high Δμ setting (-0.56 eV), Cu atoms solely replace Fe atoms with the largest atomic volume at the GB, resulting in an equilibrium Cu concentration at GB ($X_{Cu}^{gb}$) of 0.1. As Δμ decreases, Cu atoms begin to occupy additional sites, leading to a decrease in GB energy from 1.47 to 1.12 J/m$^2$, accompanied by $X_{Cu}^{gb}$ varying from 0.1 to 0.72. However, when Δμ falls below -0.78 eV, Cu atoms permeate the grain matrix, causing an increase in the system's total energy with higher Cu content. Additionally, the shape of the Cu precipitate becomes irregular, indicating an oversaturation of Cu content. Consequently, the doped structures exhibiting minimal GB energy throughout this process are selected as the doped structures for further simulations. Following a 1ns incubation at 10K, the doped structures are prepared for subsequent analyses.

The preparation of pure Fe samples involves a series of 3 steps. Firstly, the GB configuration is manually manipulated by tilting or twisting the crystals along a specific crystallographic axis. Next, the γ-surface is calculated to determine the configuration with the lowest GB energy. Subsequently, the structure is relaxed at a temperature of 10 K for a duration of 1 ns. For the doped samples, two additional steps are performed. Firstly, an MCMD routine is carried out with Δμ values ranging from -0.56 eV to -0.80 eV to identify the composition that results in the lowest energy configuration for the doped GB. Secondly, the structure is relaxed at 10 K for 1 ns. The nomenclature of these structures follows a specific format, wherein the composition, Σ value, and tilt/twist information are encoded. For instance, "Pure-Σ5-Tilt" refers to the pure Fe structure with Σ5<100>{130} symmetric tilt GBs, while "Doped-Σ13-Twist" represents the Cu-doped structure with Σ13{100} twist GBs.

### 2.2. Cascade simulation design

Following the sample preparation, collision cascade simulations are conducted using the LAMMPS platform with consistent interatomic potential settings. The design of the cascade simulations is motivated by prior studies that have highlighted the distinct defect production behavior near GBs compared to the perfect structure [21, 22, 27, 67]. Moreover, the degree of overlap between the collision cascade and the GB has been found to influence primary defect behaviors [2, 6, 24, 28]. Consequently, the present study focuses on systematically varying the distance between the primary knock-on atom (PKA) and the GB, referred to as PKA offset. To ensure the clarity and relevance of the results, certain cascade factors, such as energy and temperature, are constrained. High PKA energy introduces subcascade effects [68, 69], increases in-cascade recombination rates [68, 70, 71], leads to radiation-induced segregation (RIS) [72], and forms larger "quasi-liquid" zones [73], all of which complicate the analysis of primary defect behaviors. Considering the practical implications of understanding primary defect behavior at low energies, the PKA energy of the Fe atom in this study is set at 1 keV. Additionally, computational models predicting the neutron spectrum in fusion reactor designs indicate a high neutron flux range of 1 keV to 100 keV on reactor components [74], further justifying the selection of low PKA energy. Consistent with this approach, the collision cascade simulations are performed at a temperature of 10K. This choice serves two purposes: it limits local diffusion in the simulations and matches the conditions of helium-cooled components. Secondly, the primary focus of this study is on understanding the creation of primary damage independently of other diffusive processes, such as thermal recombination and defect annealing. Therefore, a temperature of 10K is chosen, while the exploration of diffusion effects at different temperatures is reserved for future investigations.

The collision cascade simulations are conducted according to the following detailed procedures:

a) For each cascade test, a primary knock-on atom is randomly selected from one of the atomic layers parallel to the GB plane. The distance between the PKA source layer and the GB, ranging from 0.5 nm to 6 nm with an increment of 0.5 nm, determines the degree of overlap between the collision cascade (CC) and the GB and is referred to as PKA offset.

b) Each test is initiated by accelerating the PKA to 1 keV perpendicular to the GB.

c) Following the initiation of the cascade, a 20 ps period of free evolution is performed, during which the system’s volume and energy are conserved to ensure a micro-canonical ensemble. This allows the structure to gradually reach thermal equilibrium. The defects remaining in the structure after this 20 ps evolution period are considered the primary defects.

d) To establish statistical significance, 20 independent tests are performed at each PKA offset.

e) The primary damage data, including the number and distribution of interstitials and vacancies, are collected using the Wigner-Seitz cell defect analysis in the OVITO platform [75]. The defects in the GB region and the grain matrix are counted separately to construct the defect distribution.

By following these procedures, the simulation results provide valuable insights into the primary defect behavior, particularly in terms of interstitials and vacancies, and allow for a detailed analysis of their distribution both in the GB region and the grain matrix.

During the simulation, it was observed that the PKA offset alone may not consistently indicate the degree of overlap between the collision cascade and the grain boundary. Snapshots from different structures showed variations in the maximum overlap between CC and GB at different PKA offsets. However, further analysis of the cascade snapshots revealed that the PKA, when oriented along a specific crystallographic direction, consistently develops into the CC at a specific distance relative to the pristine PKA location. To account for consistency in CC-GB interaction, a new parameter called the cascade offset was introduced. The cascade offset represents the distance from the center of the collision cascade to the GB. To determine the cascade offset, the center of the collision cascade was calculated as the mean position of all atoms with atomic displacement greater than 0.4 Å at the peak of CC evolution, specifically 0.2 ps after the initiation of the 1 keV PKA. It was found that the PKA offset effectively determines the position of the CC relative to the GB, while the cascade offset provides a more consistent indicator of the CC-GB overlapping scenarios. The correlation between the PKA offset and the cascade offset was plotted in Sup.4, revealing a linear relationship with a certain shift to the right depending on the crystallographic orientation. By incorporating the cascade offset in the analysis, a more reliable and consistent comparison among different structures regarding the CC-GB overlapping can be made. This allows for a more accurate assessment of the primary defect production behaviors and facilitates the identification of commonalities and differences among various structures.

## 3. Energetics of Grain Boundary

To investigate the influence of grain boundary on primary defect production behaviors, an analysis of GB energetics related to defect behaviors[2, 24, 26] is conducted. Specifically, the GB energy, GB vacancy formation energy, and GB interstitial formation energy are examined. The GB energy represents the stability of the GB interface, providing insights into its structural integrity. The GB energy of each tested configuration is depicted by the bars in Figure 4a. The formation energy of vacancies and interstitials within the GB signify the net energy required to create a defect and is intricately linked to the defect segregation energy denoting the driving force for defects to segregate from the matrix towards the GB. In the context of collision cascade simulations, when excess energy is introduced into the local structure by a PKA, both GB energy and defect formation energy need to be taken into consideration to comprehend the resulting damage state accurately. The calculation of GB energetics considering the effect of solute elements is performed as follows:

Eq.1. $$\gamma = \frac{E_{total} - N_{Fe} \times E_{c-Fe} - N_{Cu} \times E_{c-Cu}}{2 \times S_{gb}}$$

Eq.2. $$E_{f-i} = E_i - (E_0 + E_a)$$

Eq.3. $$E_{f-v} = E_v - (E_0 - E_a)$$

where $\gamma$, $E_{f-i}$, and $E_{f-v}$ represent the GB energy, interstitial formation energy and the vacancy formation energy, respectively. In Eq.1, $S_{gb}$ is the area of the grain boundary plane, $E_{total}$ is the total energy of the final structure, $N_{Fe}$ and $N_{Cu}$ are the number of Fe atoms and the number of Cu atoms in the structure, the $E_{c-Fe}$ is the cohesive energy of the Fe atom in pure BCC Fe structure, and the $E_{c\text{-}Cu}$ is the cohesive energy of the Cu atom in pure FCC Cu structure. In Eq.2 and Eq.3, $E_i$ (or $E_v$) is the total energy of the structure with one interstitial (or one vacancy), $E_0$ is the total energy of the pristine structure, and $E_a$ is the cohesive energy of the element in its perfect lattice, i.e., for Fe defects, $E_a = E_{c-Fe}$, and for Cu defects, $E_a = E_{c-Cu}$.

The GB energies obtained in this study fall within the range reported in previous studies [76-78], indicating consistency with existing literature. To ensure statistical significance, the GB defect formation energy calculations are performed 400 times at random sites within the GB region. The resulting distributions of interstitial formation energies ($E_{f-i}$) are presented in Supplementary Figures 2 and 3. Notably, tilt GBs exhibit distributions that closely resemble normal distributions, while twist GBs demonstrate more complex distributions with multiple peaks. The average defect formation energies for all tested GBs are summarized in Figure 4b, along with their corresponding standard deviations. In this bar chart, the defect formation energies for pure Fe GB structures are depicted in black and grey, while the energies for Cu-doped structures are shown in red and pink. The black and red bars represent the vacancy formation energy ($E_{f-v}$), and the grey and pink bars stacked on top represent the interstitial formation energy ($E_{f-i}$). The total height of each bar corresponds to the numerical summation of $E_{f-v}$ + $E_{f-i}$ for the specific GB, providing an approximation of the Frenkel pair (FP) formation energy ($E_{f-FP}$). Additionally, the defect formation energies in the pure BCC Fe lattice are presented in the rightmost bar of Figure 4b. The reported values closely align with a previous study [79] that utilized the same interatomic potential employed in the present investigation. This concurrence further strengthens the reliability and consistency of the obtained defect formation energies.

The above calculations of GB energetics reveal that the defect formation energy in the Fe BCC lattice is inherently higher compared to any of the tested GB regions. However, both the grain boundary energy (γ) and the Frenkel pair formation energy ($E_{f-FP}$) exhibit significant variations among different GB configurations (detailed values are presented in Table 1). Notably, the changes in γ and $E_{f-FP}$ are not consistently correlated across different GB configurations. For instance, consider the pure Fe symmetric tilt GBs with Σ5, Σ9, Σ11, and Σ13 values. The corresponding grain boundary energies are 1.18, 0.98, 1.25, and 0.80 J/m$^2$, respectively. However, the Frenkel pair formation energies for these GB structures are 2.56, 2.15, 2.33, and 2.49 eV, respectively. There is no apparent dependence between γ and $E_{f-FP}$ observed within the tested GB configurations, indicating that these two properties are independent of each other for a given GB configuration. Reports from previous studies [35, 80-82] suggest that defect formation energies in GBs are influenced by local atomic structures, such as the site volume within the GB. Furthermore, the grain boundary energy may be more dependent on surface energy and the dihedral angle between the two planes [34, 83, 84], as described by Mullins equation: $\frac{\gamma_{GB}}{\gamma_s} = 2\cos\left(\frac{\Psi}{2}\right)$, where Ψ represents the dihedral angle. These factors contribute to the complexity of GB properties and highlight the need to consider multiple parameters when studying GB behavior.

It is also worth noting that, in our dataset, the prepared Σ5 twist GB exhibits a unique characteristic of negative interstitial formation energy at the GB, both in the pure and doped cases. This suggests that the structure will be more energetically stable with the presence of extra atoms in the GB region. Additionally, it is observed that the doped Σ5 twist GB has a higher Frenkel pair formation energy $E_{f-FP}$ compared to its pure Fe counterpart, while the dopant reduces the $E_{f-FP}$ in other tested GB configurations. Overall, the doping effects on the energetics (γ and $E_{f-FP}$) show inconsistency among all the tested structures. This highlights the need for separate investigations into grain boundary energy and defect formation energy to gain a comprehensive understanding of their respective effects on cascade-induced defect production.

## 4. Grain Boundary Defect Production

To gain insight into the influence of grain boundary on defect production, an examination of defect behaviors at various distances from the GB is essential. In this regard, a series of snapshots was obtained during the collision cascade simulation to depict the evolution of the cascade near the GB. Figure 5a illustrates one such snapshot taken at the peak phase of a developing cascade, with a cascade offset of 2 nm relative to the pure Fe Σ5 twist GB, as defined earlier. In this representation, atoms are color-coded based on their displacement magnitude from their initial positions, while the vectors indicate the corresponding direction. To highlight the most active part of the cascade, atoms with a displacement below 0.4 Å were omitted from the image. The observed cascade in this snapshot exhibits a symmetrical, spherical shape, with the center of the cascade offset from the original position of the PKA. Despite the reach of several focused chains to the vicinity of the GB due to the 2 nm cascade offset, there appears to be a lack of significant interaction between the cascade core and the GB, as indicated by the cascade morphology. This snapshot provides valuable insights into the initial behavior of the collision cascade in the vicinity of the GB, further analysis of cascade snapshots at different cascade offsets will contribute to a more comprehensive understanding of how the presence of the GB influences the production of defects.

Figure 5b provides insights into the effects of the GB on the defect production process by comparing two distinct defect evolutions at different cascade offsets. The lower trend depicts the defect evolution when the cascade develops 2 nm above the GB (2nm cascade offset), while the upper trend represents the defect evolution when the cascade overlaps with the GB (0nm cascade offset). Insets in Figure 5b show snapshots at different stages of the cascade based on the models described in Ref. [85]:

a. <0.2 ps (inset I-II): This phase corresponds to the supersonic collisional phase, characterized by a significant number of displacement defects and focused collision chains (long directional displacement sequences along closest packed direction) generated by the collisions. The size of the cascade, shaped by energetically activated atoms with significant displacements, increases over time and reaches its maximum at around 0.2 ps, which is considered the peak of the cascade.

b. 0.2 ~ 10 ps (insets III): This phase corresponds to the in-cascade recombination phase, where the size of the cascade gradually shrinks. Recombination takes place near the cascade center, resulting in a decrease in the population of Frenkel pairs. Most of the focused chains zip back completely, but some may evolve into replacement sequences, leading to the generation of remote interstitials and split Frenkel pairs.

c. 10 ps ~ 100 ps (inset IV): This phase corresponds to the primary damage phase, where the kinetic evolution of the cascade comes to a halt. The surviving defects begin to evolve at a relatively slow rate due to thermal vibration.

Since the focus of the present study is on the primary damage, it is important to consider the structural components that emerge during the cascade evolution. The interaction between the cascade and the GB may influence the detailed features of the primary damage state, such as the population and distribution of the survival defects at the conclusion of the kinetic evolution of the cascade.

The inset IV of Figure 5b presents snapshots of the primary damage stages, specifically highlighting the peak damage region (PDR) encompassed by the yellow displacement vectors near the center of cascade. The PDR, serving as the core of the cascade, contains a significant portion of the surviving vacancies within its range of 1nm in radius. Comparing the shapes of the PDRs in the two scenarios, at 0 and 2 nm cascade offsets, reveals a notable contraction of the PDR within the GB when overlap occurs. Additionally, the overall number of survival Frenkel pairs generated by the cascade, counting defects both in the GB and in the matrix, is significantly increased in the overlap scenario. These observations underscore the pronounced attraction effect exerted by the GB on the cascade and survival defects. Previous research has discussed two processes that may account for this phenomenon of defect attraction. The first process [24, 28, 30, 86-88] relates to the cascade-GB interaction, where the presence of the GB impedes the passage of mass and energy from the cascade, thereby causing damage to the GB through energetic cascade-GB interactions. The second process [6, 22, 27, 29, 86] involves the defect-GB interaction, whereby nearby defects are drawn towards the GB due to thermodynamically favorable energetics in the GB region. These studies shed light on the essential role played by both GB energetics, encompassing GB energy and defect formation energy, and GB structural features, including atomic density and local structure, in influencing the primary defect production within collision cascades.

The subsequent discussion on the factors influencing defect production necessitates an analysis of defect production statistics under different conditions, which are summarized in Figure 6. The plot depicts the total number of Frenkel pairs, encompassing GB defects, generated at the end of the simulation, with the cascade offset serving as the organizing parameter to account for the extent of cascade-GB interaction. The left panel focuses on defect production trends near tilt GBs, while the right panel illustrates the trends near twist GBs. The grey shading represents the Frenkel pair production rate within the first standard deviation in the pure Fe BCC lattice. It should be noted that the tests on the pure Fe lattice were conducted along all crystallographic orientations equivalent to the tests performed on the GB structures. In the majority of the GB structures, the results indicate a higher rate of defect production when the cascade occurs within a cascade offset range of ±1 nm. The influence of GB on defect production varies under different conditions. For instance, in the pure Fe Σ5 twist GB, the defect production rate can reach up to four times that of the matrix. However, in the Σ11 twist GB doped with Cu, the production rate can drop below the matrix value near 0nm cascade offset. The trends also suggest that the presence of dopants may decrease the overall defect production at the same cascade offset in certain GBs, such as Σ5 twist and Σ13 twist, while it may increase defect production in other GBs, such as Σ11 tilt and Σ13 tilt. These results indicate a complex nature of the interaction, highlighting the need for detailed discussions to elucidate the conditions governing the effects of dopants and GBs on defect production.

The presence of a GB has been shown in previous studies to have different effects on the spatial distribution of survival interstitials and vacancies near the GB [2, 6, 21, 22, 24-30]. It exhibits a biased attraction towards interstitials over vacancies. To gain insights into the factors influencing this attraction behavior, a detailed analysis of defect distribution under different cascade-GB interactions is necessary. The degree of interaction is quantitatively categorized into three scenarios based on the maximum size of the cascade, approximately 2nm in radius at 1keV cascade energy (Figure 5): 1. Overlap scenario, where the cascade directly overlaps with the GB (cascade offset is 0nm). 2. Intermediate scenario, where the cascade is 1nm away from the GB (cascade offset is 1nm). 3. Indirect interaction scenario, where the cascade is 2nm away from the GB (cascade offset is 2nm). The defect distributions in all tested structures under these scenarios are summarized in Figure 7. Since the number of interstitials and vacancies remains the same throughout the simulation, a direct comparison of the percentage of interstitials and vacancies can be made. Figure 7a illustrates the defect distribution trends, where red and blue denote interstitials and vacancies, respectively. Solid marks represent the pure structure, while open marks represent the doped structure. The analysis reveals the following observations:

1. The fraction of GB defects decreases as the cascade offsets increase, indicating a reduced presence of defects in the GB compared to the overall defect population.
2. In all three scenarios, a higher proportion of interstitials is observed in the GB region compared to vacancies.
3. While the fraction of GB vacancies shows a significant decline from 0 nm to 1 nm cascade offset, the fraction of GB interstitials remains relatively high.

These findings suggest a biased absorption or attraction of interstitials by the GB structure. This phenomenon is observed not only when part of the collision cascade enters the GB region (overlap or intermediate scenario) but also when only long replacement sequences can reach the GB region (indirect interaction scenario). Notably, the statistical analysis indicates that the chemical composition of the GB does not have a significant impact on the efficiency of defect attraction by the GB.

The detailed defect distributions are presented in Figure 7b, revealing the preference for interstitials over vacancies within the GB region. The colors representing interstitials exhibit a more reddish hue compared to vacancies. However, the actual fraction of GB defects varies significantly among different GB configurations:

1. At 0 nm offset, in the Σ9 twist pure Fe GB, the fraction of GB vacancies is approximately 20% lower than that of vacancies in the matrix. This means that around 60% of the vacancies are located in the GB region. In contrast, in the pure Fe Σ5 twist GB, over 80% of the vacancies reside in the GB region. With the total defect numbers in Figure 6, the results also indicate that the number of GB defects is significantly lower when the cascade develops in the Σ9 twist GB compared to the Σ5 twist GB.

2. From 0 nm offset to 1 nm, in the Σ9 twist pure Fe GB, the fraction of GB vacancies decreases from around 60% to approximately 10%, while the fraction of GB interstitials remains around 80%.
3. The fraction of GB defects is similar between the pure structures and their doped counterparts, although differences are observed in certain GB configurations.
4. The largest disparity between the fraction of GB interstitials and GB vacancies is observed at 1 nm offset.

These findings suggest two key observations:1. The preferential absorption of primary interstitials by the GB is influenced by both the characteristics of the GB itself and the presence of dopants. 2. Vacancies and interstitials can be generated not only in close proximity to each other within the same region but also at a certain distance apart. The degree of cascade-GB interaction can activate or deactivate various mechanisms of defect production. Prior studies [85, 89, 90] have discussed the two processes through which defects are generated during the evolution of collision cascades. These processes involve the displacement of lattice atoms, resulting in adjacent Frenkel pairs, as well as the formation of split Frenkel pairs through replacement collision sequences.

To determine the dominant defect process under different scenarios, an investigation of two observed mechanisms in the simulation is presented in Figure 8. Figure 8a illustrates the detailed distributions of survival interstitials generated by cascades in the BCC pure Fe single crystal (represented in grey) and cascades near the pure Fe Σ5 twist GB with cascade offsets of 1 nm (shown in blue) and 2 nm (depicted in red). The distributions are plotted based on the distance from the cascade center along the incident direction, with the x-axis origin representing the cascade center. In the case of the cascade with a 1 nm offset to the Σ5 twist GB, the interstitials are predominantly located within the GB region. In contrast, the interstitial distribution in the single crystal exhibits a broader range and is approximately symmetrical about the center. Comparing the cases with CC-GB interaction to the single crystal case reveals two significant phanomena:

1. When the cascade is as close as 1 nm to the GB, the GB not only "collects" interstitials produced by nearby cascades but also influences the cascade evolution and absorbs interstitials that would have been created in other directions.
2. For the cascade at a 2 nm offset to the GB, the defect distribution is also influenced by the GB, but the spatial distribution of interstitials above the GB closely resembles the distribution observed in the single crystal results. This suggests that at this offset, the GB may not actively participate in the defect production process but rather assists in redistributing defects by capturing nearby interstitials.

Figure 8b provides an example of displacements following a cascade with a 1 nm offset, illustrating the detailed defect production processes within and near the GB. In the figure, purple spheres represent survival vacancies, red spheres represent survival interstitials, and the small blue spheres with an ordered pattern indicate the GB region. In this particular snapshot, the vacancy (purple sphere) at the top serves as the location of the primary knock-on atom, while the center of the cascade is situated near the vacancy-rich PDR between the cascade and the GB. Notably, all survival interstitials are trapped within the GB region. The observed phenomenon at a 1 nm cascade offset can be attributed to the interaction between the shockwave and the GB. Specifically, when the shockwave enters a relatively loosely packed GB region, it reduces the pressure of the cascade and dissipate cascade energy through the GB. Consequently, the initiation of focused chains within the cascade is suppressed. This reduction in focusing activities leads to a decrease in the formation of replacement collision sequences, ultimately curbing the generation of remote interstitials associated with these RCSs. As a result, the shockwave-GB interaction inhibits the formation of remote interstitials, and some of the survival interstitials within the GB are likely formed through atomic displacements occurring within the GB itself, which accommodates the stress caused by the shockwave and absorbs a portion of the cascade energy.

To facilitate a better understanding and comparison of different mechanisms involved in GB interstitial production, the survival GB interstitials can be categorized into two distinct types, as depicted schematically in Figure 8c: 1. External GB interstitial: This type of interstitial is formed through replacement collision sequences originating from a lattice vacancy. In the snapshot, this type of GB interstitial is connected to a lattice vacancy through a long sequence of displacement vectors. 2. Internal GB interstitial: These interstitials are generated through displacements occurring within the GB itself. Importantly, this type of GB interstitial always appears in conjunction with a GB vacancy, forming a GB Frenkel pair. By differentiating between these two types of GB interstitials, insights can be gained into the underlying mechanisms responsible for their formation and better analyze their contributions to the overall defect distribution within the GB region.

The number of the GB interstitials in all tested structures at the end of the simulations are summarized in Figure 9 based on the categories discussed above. When the cascade occurs within the GB, a significant portion of the GB interstitials are the internal GB interstitials, produced by displacement within the GB. As the kinetic energy decreases from the core to the periphery of the cascade, the outer regions of the cascade may not possess sufficient energy to initiate the displacements inside GB. Consequently, the population of internal GB interstitials is anticipated to decrease with increasing cascade offset beyond 0 offset, as depicted in Figure 9a. The number of internal GB interstitials reaches a peak at 0 nm offset, subsequently decreasing as the offset increases. Conversely, the number of external GB interstitials peaks at 1 nm offset, coinciding with the proximity of the 1 nm radius peak damage region of the 1 keV cascade to the GB. This distance proves to be the most effective for focused chains to form and enter the GB region. When an energetic atom impacts the undistorted lattice within a critical angle along the closest packed directions, a focused collision chain can be formed. The energy is gradually reduced with each subsequent collision within the chain. Consequently, the probability of creating a GB interstitial diminishes as the length of the focused chain increases. At small cascade offsets where the PDR overlaps or partially overlaps with the GB, the change in lattice orientation inhibits focusing activity, resulting in a decrease in the number of external GB interstitials. Therefore, within the PDR radius (1 nm cascade offset), the number of focused chains increases as the offset increases. However, beyond the PDR radius, the number of focused chains decreases with increasing offset. This behavior contributes to the peak in the number of external GB interstitials at 1 nm offset.

## 5. Energetics of Biased GB Defect Formation

Another expected consequence of the production mechanism is that the number of external GB interstitials does not exhibit a strong correlation with GB properties, as the majority of the process occurs within the lattice. Figure 10a and b illustrate the number of external GB interstitials as a function of GB energy at 0 nm and 1 nm cascade offsets, respectively, depicted in the grey shade. It is evident that there is no significant correlation between the number of external GB interstitials and GB energy. In contrast, the production of internal GB interstitials, which is representative of GB Frenkel pair formation, seems to be influenced by two GB properties: 1. GB energy (γ), indicating the ease of GB deformation, and 2. GB Frenkel pair formation energy ($E_{FP}^{GB}$), which pertains to the energy required for defect generation. GB configurations with higher GB energy are more prone to deformation than those with lower GB energy. Additionally, higher $E_{FP}^{GB}$ values suggest a greater energy cost for GB Frenkel pair formation. Consequently, it is reasonable to anticipate an increased GB defect production rate with higher γ values and lower $E_{FP}^{GB}$ values.

The correlation between defect production and GB energy (γ) is depicted in Figure 10, where the colored marks represent the number of internal GB interstitials produce after each cascade. Overall, the trends indicate a certain degree of correlation between internal GB interstitial production and GB energy, as anticipated. In general, as GB energy increases, so does the production of internal GB interstitials. However, when the results from twist GB structures are excluded from the plots (triangles in Figure 10a and b), the data points from tilt GB structures (squares) do not exhibit a significant correlation with GB energy. Figure 10c and d provide a detailed comparison after grouping the results into tilt GBs (grey shades) and twist GBs (pink shades). The internal GB interstitial production in different structures is shown at 0 nm offset (c) and 1 nm offset (d), with the results of pure GB and doped GB structures presented side by side. The corresponding GB energies are listed below the bars. In the structures represented by the pink shades in Figure 10c and d (structures with twist GBs), the number of internal GB interstitials is lower in the doped structures compared to their pure Fe counterparts at both cascade offset cases. However, the effect of doping is not consistent across all tested structures, and the structures with tilt GBs (depicted in the grey shades) exhibit a more complex pattern due to a more complicated GB configuration.

As previously discussed, the production of internal GB interstitials may also be influenced by the energy cost of GB Frenkel pair formation, represented by $E_{FP}^{GB}$. Additionally, the dopant can have different effects on two important properties: γ and $E_{FP}^{GB}$. On one hand, the dopant may help decrease γ, making the GB more stable. On the other hand, the dopant may also reduce $E_{FP}^{GB}$, thereby lowering the energy required to form a GB Frenkel pair. The former effect would decrease the defect production, while the latter would promote it, thus counteracting the former effect. Figure 11a illustrates the doping effects on both properties in the tested structures. It shows the relative change in properties from the pure Fe GB to the doped GB ($\left(E_d - E_p\right)/E_p$, where Ed represents the energetics of the doped structure and Ep represents that of the pristine structure). The percent changes in γ are depicted as green bars, while the changes in $E_{FP}^{GB}$ are shown as red bars. For the tested tilt GBs, the doping process increases GB energy but significantly decreases the defect formation energy. However, the doping effect on $E_{FP}^{GB}$ in the tested twist GBs is relatively small. Considering both properties together, the internal GB interstitial production was analyzed as a function of the combined parameter $\gamma/_{E_{FP}^{GB}}$ and depicted in Figure 11b. The correlation coefficient between the defect production and the combined parameter showed improvement from 0.817 (production vs. γ) and -0.822 (production vs. $E_{FP}^{GB}$) to 0.895 (production vs. combined parameter). Previous studies [80-82] have indicated that the GB defect formation energy can be significantly influenced by the local configuration. Therefore, the enhanced correlation with the combined parameter suggests a strong correlation between the GB defect production and both factors.

In conclusion, the defect production processes during low-energy cascades can be elucidated based on the spatial relationship between the cascade and the GB:

Firstly, when the GB is situated within the PDR of the cascade, the defect production process is primarily governed by the multi-factor GB Frenkel pair production mechanism. This process is characterized by the combined parameter $\gamma/_{E_{FP}^{GB}}$, resulting in a relatively small number of external GB interstitials and lattice interstitials introduced by focusing events (RCS). As the cascade offset increases, the fraction of RCS-induced interstitials rises while the fraction of GB Frenkel pair production diminishes.

Secondly, when the GB lies beyond the PDR but within the shockwave radius (2nm) of the cascade, the defect production process is dominated by RCS-induced interstitials, which are independent of GB properties. The production of GB Frenkel pairs is comparatively limited, rendering GB properties less influential on the defect production process. Increasing the cascade offset in this regime leads to a decrease in both external and internal GB interstitials, as demonstrated by Figure 9 in the range of 1nm to 2nm cascade offset.

Lastly, when the GB is situated beyond the shockwave radius of the cascade, the total number of defects generated in the structure is not significantly influenced by the presence of the GB. However, the GB acts as a collector for RCS-induced interstitials within its capture radius. Further investigation would be needed for a more detailed frame, and a more general range of cascade energies and GB configurations.

Furthermore, in addition to the number of defects, considering the energy associated with defect formation provides another perspective on cascade damage. It is important to recognize that the energy required to form the same number of defects can vary significantly among different structures due to variations in defect formation energy. For instance, as demonstrated in Figure 4, the Σ5 twist GB, both in its pure form and when doped, exhibits a negative GB interstitial formation energy, resulting in a lower Frenkel pair formation energy within the GB compared to the bulk matrix. With these considerations in mind, the potential energy change before and after the collision cascade, referred to as the stored energy, has been calculated and compared in Figure 12. This concept is akin to the Wigner energy commonly studied in irradiated graphite [91-95]. The stored energy was determined by subtracting the potential energy of the pristine structure from that of the damaged structure at a temperature of 10K. Figure 12a presents the stored energy information at both 0nm and 2nm cascade offsets for all the tested structures, including the single crystal BCC structures with the PKA direction aligned along the corresponding crystallographic orientation.

Despite the significantly larger number of defects observed in structures with GBs compared to single crystal structures, the stored energy in the single crystal counterparts surpasses that of the doped and undoped structures with GBs. The impact of doping remains inconsistent across different structures, with a general trend of reduced stored energy in structures featuring tilt

GBs at both cascade offsets. However, this trend does not hold consistently for structures with twist GBs. Notably, two structures, namely the pure Fe Σ5 twist GB and the Σ9 twist pure Fe GB, exhibit exceptionally low stored energy values at 0 nm cascade offset. The former can be attributed to the negative GB interstitial formation energy, while the latter is a consequence of the cascade-induced GB reconfiguration process, which will be elaborated upon in a subsequent section.

When considering the entire range of tested structures, Figure 12b portrays the residual cascade damage in terms of stored energy at various cascade offsets. It is evident that the stored energy increases with increasing cascade offset, signifying that cascades developed in close proximity to GBs inflict less damage in terms of potential energy with a higher portion of cascade energy dissipated. Moreover, the impact of doping on the stored energy remains inconsistent across different structures, mirroring the variability observed in the number of defects. These findings underscore the intricate nature of defect production near GBs, which is influenced by a multitude of factors encompassing both energetics and structural characteristics.

## 6. Grain boundary structural motifs underpinning displacement behaviors

### 6.1. Grain boundary structure and grain boundary reaction

As forementioned, within the tested GB structures, one particular configuration exhibits a unique behavior when subjected to a nearby developing cascade. Specifically, the pure Fe Σ9 twist GB, which possesses the lowest GB energy among the tested GBs, undergoes a configuration change during cascade evolution. Initial investigations suggest that the Σ9 twist GB may transform into two Σ3 twist GBs. Although the Σ9 twist GB is typically stable, requiring a net input of energy to increase or decrease the misorientation angle [96], the grain boundary reaction can be activated by relatively low-energy collision cascades. Detailed snapshots of the structure in this study are presented in Figure 13, wherein atoms are color-coded based on atomic volume: (a) parallel view of the GB before and after the cascade, (b) zoomed-in view highlighting the closest packed direction with indicated arrows, and (c) vertical view to the GB plane.

Examination of figures (b) and (c) reveals that the initial Σ9 twist GB configuration possesses two strained layers with higher atomic volume compared to the matrix. Subsequent cascade-induced atom reconfiguration results in a more evenly distributed three-layer structure with reduced strain. Notably, the top view in figure (c) demonstrates a change in the entire coincidence site lattice pattern after the cascade. Considering the pattern, energy reduction following the transition, and the misorientation angles between layers, this structural transition is likely a GB disconnection process leading to the transformation of the Σ9{110} twist GB into two Σ3{110} twist GBs [97]. The exceptionally low stored energy observed in the Σ9 twist GB after the cascade (shown in Figure 12) is indicative of the benefits derived from this exothermic process. Among the tests conducted on the pure Σ9 twist GB structure, this transition was observed in 70% of cases near a 0 nm cascade offset, while 0 occurrence was observed in the doped structure at the same offset. This disparity suggests that the presence of the dopant stabilizes the Σ9 twist GB structure and inhibits the transformation process.

It is pertinent to highlight the notable characteristics of the Σ9 {110} twist GB, which possesses a misorientation angle of 38.94º, while the twist Σ3 {110} GB exhibits a misorientation angle of 70.53º. Consequently, the misorientation between two primary grains separated by two twist Σ3 GBs precisely amounts to 38.94º. This implies that a minimal adjustment of one atomic layer within the grain boundary region is required to complete the observed transition from a single Σ9 twist GB to two twist Σ3 GBs, without necessitating any rotational or translational displacements in the grain matrix. The activation energy associated with this transition is expected to be relatively low, as a partial rearrangement of a single atomic layer suffices to complete the transition. Moreover, the transition itself is exothermic, thereby supplying the requisite energy for subsequent processes. In contrast, the process of grain boundary migration entails the rearrangement of at least two atomic layers. It is noteworthy that certain grain boundary reactions may possess lower activation energies for transforming into alternative grain boundaries compared to undergoing grain boundary migration. Consequently, the reconfiguration of high-angle grain boundaries (HAGBs) through disconnection events has the potential to occur in real materials. However, the ability of such reconfigurations to transpire under specific conditions is contingent upon the characteristics of the grain boundary structure. Notably, the observed reconfiguration from a Σ9 {110} twist GB to two twist Σ3 GBs is not observed in the structure featuring the Σ11{110} twist GB, despite the latter exhibiting similar energetic properties to the former.

### 6.2. Grain boundary structure and focusing activities

The structural characteristics of grain boundaries can also exert influence on the evolution of cascades. Figure 9a reveals that the cascade offset has a comparatively diminished impact on the internal GB interstitial production in the Σ9 twist and Σ11 twist GBs compared to other tested GBs. This suggests that the presence of Σ9 twist or Σ11 twist GBs in the vicinity of a cascade does not significantly influence the cascade's evolution in the same manner as other GBs. Figure 14 provides a visual representation of three distinct cascade evolutions: within the pure Fe Σ5 twist GB, the pure Fe Σ11 twist GB, and the doped Σ11 twist GB. In Figure 14a, a vertical view of the GB plane reveals an inhomogeneous distribution of atoms on the Σ5 twist GB plane, whereas the atoms on the Σ11 twist GB plane exhibit an aligned and lattice-like arrangement. This orderly atomic arrangement within the Σ11 twist GB facilitates the presence of preferential directions for the focusing effect [98-102], thereby enabling the initiation of focused chains emanating from the cascade. Figure 14b demonstrates that the cascade on the pure Fe Σ5 twist GB plane merely generate collision sequences, whereas the cascade on the Σ11 twist GB plane comprises several elongated displacement sequences along the preferential direction, as focused chains. Both the doped and undoped Σ11 twist GBs exhibit the potential for initiating focused chains, with the Cu dopant having minimal distortion on that Σ11 twist GB configuration. These focused chains possess the ability to zip back without leaving behind any defects. When the leading edge of a focused chain becomes trapped, a split Frenkel pair is created, constituting RCS event. A similar definition is employed in Becquart's work [89], whereby RCS events lead to the production of defects, while focused chains themselves do not. However, both sequences dissipate energy with each collision.

In the Σ5 twist GB configuration, the presence of discontinuous closest packed directions of atoms restricts the occurrence of focusing activities. Conversely, in the Σ9{110} twist GB or Σ11{110} twist GB configurations, the maintenance of closest packed directions within the GB plane facilitates the occurrence of focusing activities. This enhanced focusing activity within these GBs plays a vital role in dissipating energy from the cascade. Consequently, the Σ11 twist GB plane (and the Σ9 twist GB) exhibits a reduced ability to generate defects compared to configurations lacking a well-organized and uniformly distributed GB plane. This phenomenon elucidates the smaller fraction of GB defects observed in these two GB configurations, as depicted in Figure 7b.

To quantitatively evaluate the capacity of a GB plane to facilitate focusing activities, the tortuosity of the GB plane is calculated. The tortuosity is defined as the ratio of the shortest path connecting the nearest atoms to the straight-line distance between two points located approximately 5nm apart. For example, the tortuosity of the GB plane consisting the closest packed direction will be 1. Figure 15a illustrates the number of focused chains at the peak of cascade evolution with a cascade offset of 0nm, plotted against the tortuosity of the GB plane. Focused chains are identified using a criterion of a minimum length of 2nm. In general, an increase in tortuosity corresponds to a decrease in the number of focused chains. Notably, the significantly low number of focused chains observed in the doped Σ5 twist GB (indicated by blue open triangles) may be attributed to the low atomic density in the GB region (92.6% of the density in the matrix, the lowest among all tested structures). This factor may further impede the formation of focused chains along other directions by reducing the pressure in the cascade core.

In Figure 15b, the internal GB interstitial production is categorized into two groups based on the number of focused chains observed. The first group comprises structures that develop 4 to 8 focused chains at the peak of the cascade, while the second group consists of structures that generate 9 to 13 focused chains. Interestingly, the group with a higher number of focused chains exhibits a lower production of internal GB interstitials. Owing to the distinct characteristics of cascade evolution in these groups, the distributions of GB defects also differ noticeably. For instance, in Figure 15c, the survival vacancies and interstitials in the pure Fe Σ5 twist GB (category I) tend to be closely situated within a smaller region, with vacancies near the center and interstitials in the periphery. The distribution in this case exhibits relative symmetry. Conversely, the survival GB defects in the pure Fe Σ11 twist GB (category II) are more dispersed and lack symmetry.

Hence, in addition to the influence of GB energetics on cascade evolution, the structural characteristics of GBs play a crucial role in the production of defects during cascades and can offer insights into predicting the defect generation process. To gain a more comprehensive understanding, it is essential to conduct further investigations encompassing a wider range of GB configurations, varying PKA energy spectra, and incorporating realistic distributions of GB configurations. Such endeavors will enable the exploration of weighted average effects of GBs and dopants on cascade-induced damage, leading to more informative and practical conclusions.

## 7. Conclusions

In this study, a methodical examination of the complex mechanisms governing the generation of primary defects within doped iron grain boundaries subjected to low-energy collision cascades was conducted through molecular dynamics simulations. The introduction of sophisticated analytical methodologies and techniques facilitated the elucidation of the grain boundary (GB) properties' paramount role in dictating the formation of radiation-induced defects. The findings demonstrate that:

i. The interplay between the collision cascade and the grain boundary invariably engenders a proliferation of defects exceeding those observed within a flawless lattice devoid of a GB. Nonetheless, this augmentative effect is markedly contingent upon the specific GB configuration. In the instance of the Twist $\Sigma5\{100\}$ GB, the incidence of Frenkel pairs surpasses that within the pristine lattice by a factor exceeding threefold, whereas in the Twist $\Sigma9\{110\}$ and $\Sigma11\{110\}$ GBs, the defect generation is not significantly influenced by the collision cascade-GB interaction.

ii. A pronounced bias for the absorption of interstitials over vacancies is exhibited by the grain boundary. The presence of a GB alters the evolution of the cascade and the spatial distribution of interstitials by pulling in the replacement sequences. Within a cascade offset of 1 nm, the GB distinctly favors interstitial absorption, with an average up to 90% of interstitial sites located within the GB, in stark contrast to a less than 40% average for GB vacancies.

iii. Upon categorizing the grain boundary interstitials according to their source of origin, it is observed that the GB energy exerts minimal influence on the quantity of external GB interstitials, which are translocated from the lattice into the GB by the replacement sequences. Conversely, a lower GB energy is associated with a diminished production of internal GB interstitials, generated through displacements within the GB itself. This trend is evident in both 0 nm and 1 nm cascade offset scenarios, with greater significance in the 0 nm offset case, where cascades evolve directly within the grain boundaries.

iv. A negative correlation exists between defect production and the GB Frenkel pair formation energy, with a correlation efficiency comparable to that with GB energy. However, the impact of doping on the GB energy and GB Frenkel pair formation energy varies across different GB configurations. A combined parameter ($^{\gamma}/_{E_{FP}^{GB}}$) exhibits a more robust correlation with defect quantity, suggesting a multifaceted dependency of defect production on various factors.

v. Beyond GB energetics, the geometric characteristics of the GB also influence cascade evolution and defect production. A noteworthy observation is the cascade-induced reconfiguration of the entire GB structure, as seen in samples with a Twist Σ9 pure Fe GB. Moreover, the displacement sequence behavior is modulated by GB geometry, with a smoother GB plane promoting the formation of extended displacement sequences, whereas a more irregular GB plane restricts the cascade to a comparatively limited area. These phenomena potentially impact the number of defects generated by altering the routine of energy dissipation.

Copper doping, in particular, was shown to modify GB energies and defect formation energies, highlighting the nuanced relationship between GB characteristics, dopant elements, and defect dynamics. The implications of these findings extend to the design and development of materials with enhanced radiation tolerance, particularly for applications in nuclear reactors and other radiation-prone environments. By tailoring GB properties and selectively introducing dopant elements, it is conceivable to engineer materials that exhibit superior resistance to radiation-induced damage. Further work, expanding on the range of dopant elements, GB configurations, and energy cascades, is essential to fully harness the potential of GB engineering in developing materials capable of withstanding the challenges posed by radiation exposure.

## Acknowledgements

BPU was supported by FUTURE (Fundamental Understanding of Transport Under Reactor Extremes), an Energy Frontier Research Center funded by the U.S. Department of Energy, Office of Science, Basic Energy Sciences. Los Alamos National Laboratory, an affirmative action equal opportunity employer, is managed by Triad National Security, LLC for the U.S. Department of Energy's NNSA, under contract 89233218CNA000001.

**Figures**

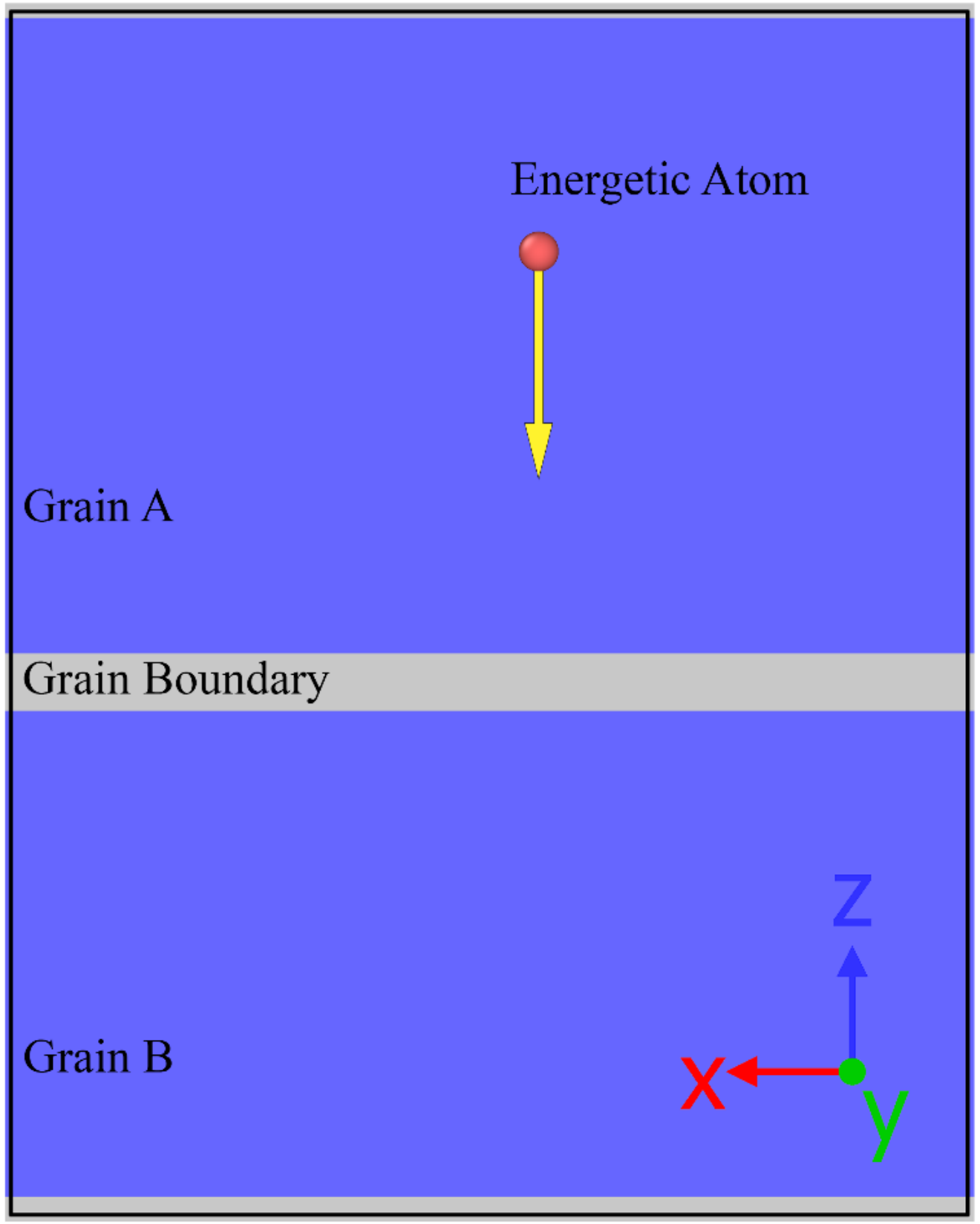

Figure 1 Schematics of the initial state of the collision cascade simulation.

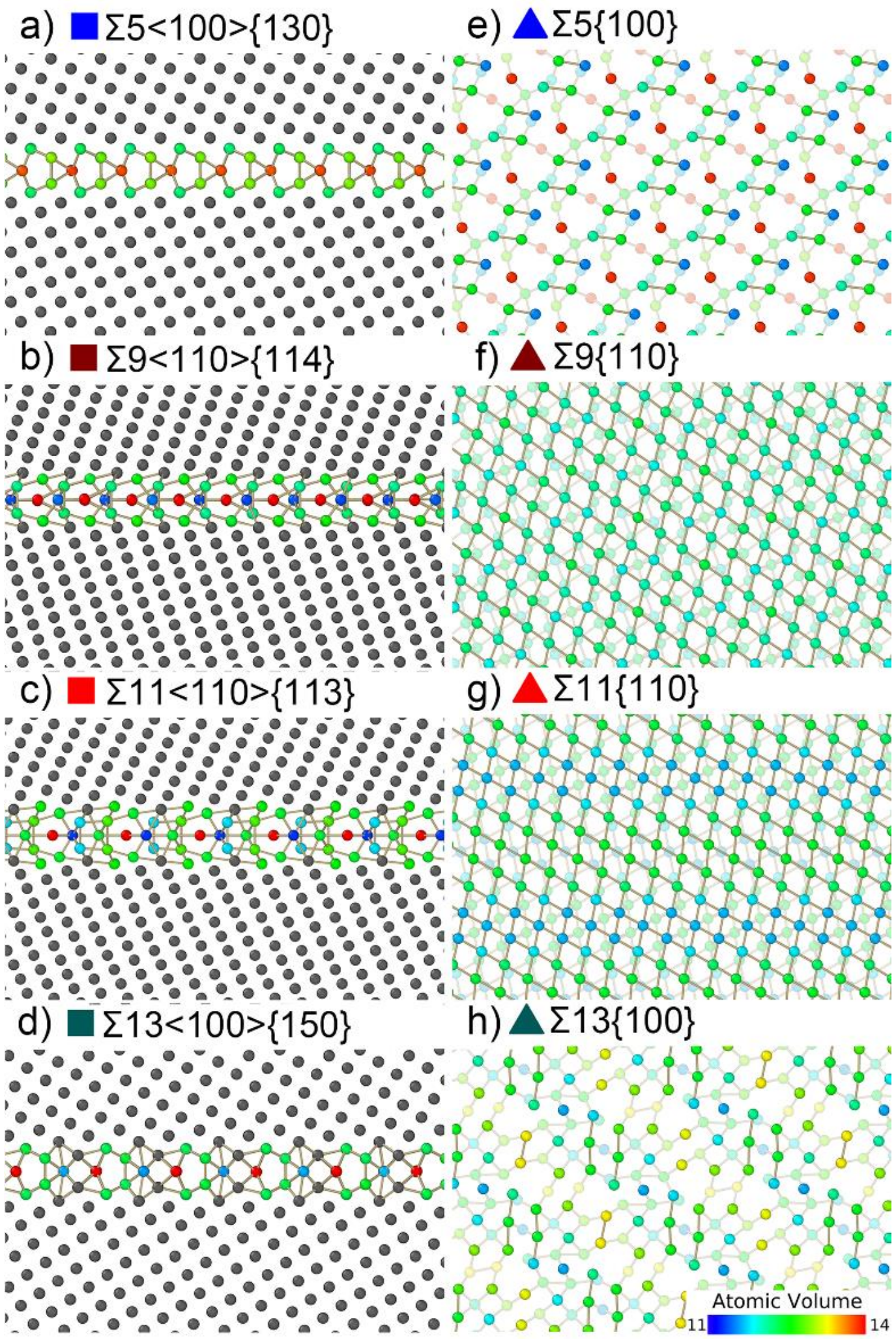

Figure 2: A collection of structures with different grain boundary configurations. To test the effect of doping on the response of the grain boundaries to collision cascades, both pure iron grain boundaries and grain boundaries fully doped with copper have been tested under the same settings of simulation experiments. Four symmetric tilt (a-d) and twist (e-h) grain boundaries have been examined in this work. The colored symbol next to each label identifies that boundary in the subsequent figures. The colors of the atoms represent the local atomic volume around that atom, highlighting the variations in volume at the grain boundaries themselves. The tilt boundaries are viewed down the tilt axis while the twist boundaries are viewed normal to the twist axis.

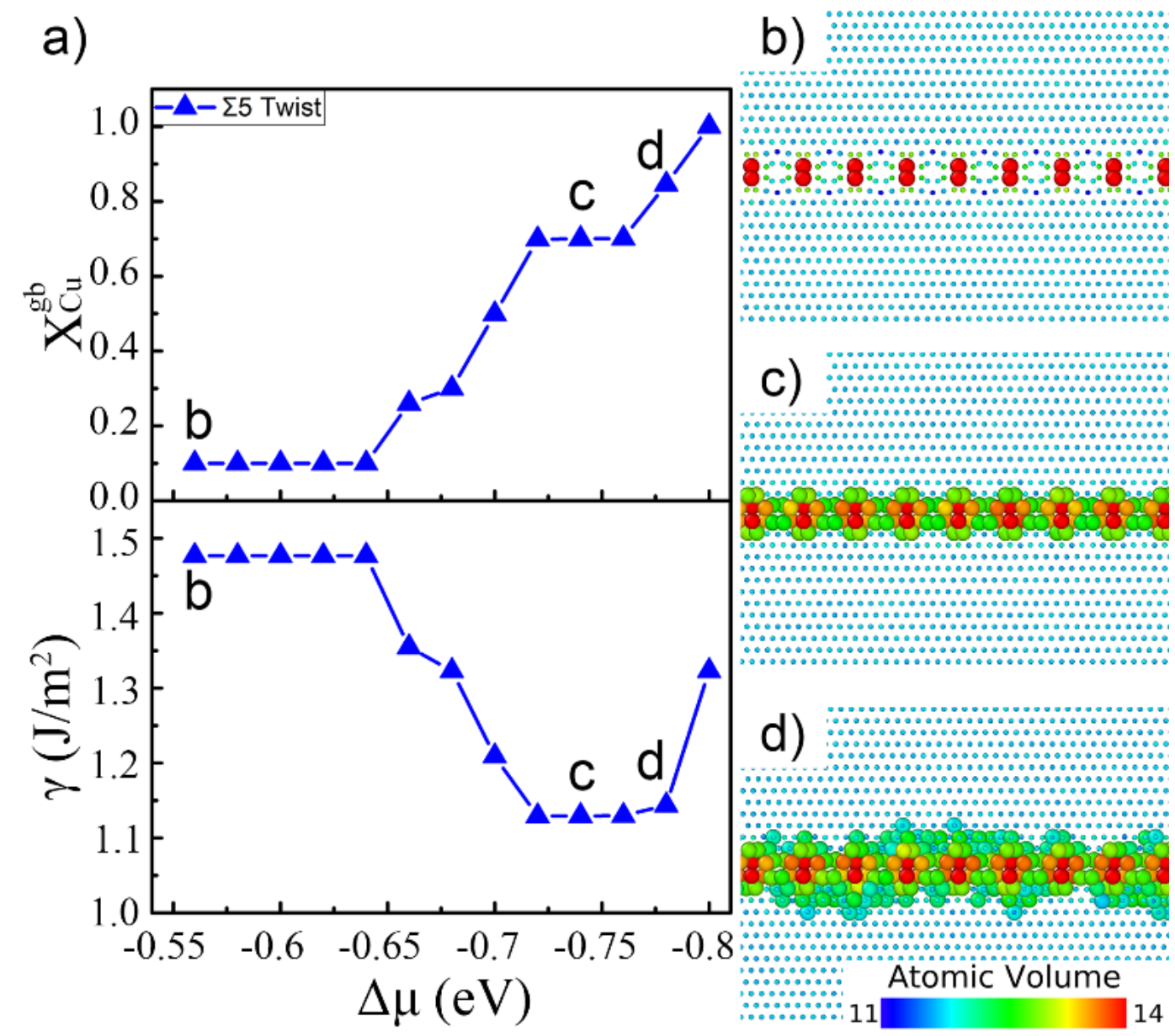

Figure 3: The variation in both grain boundary concentration of Cu ($X_{Cu}^{gb}$) and the resulting grain boundary energy (γ) as Cu is introduced into one representative grain boundary, the Σ5 twist boundary. By varying the chemical potential (Δμ) used by the VCSGC-MC routines in LAMMPS, the Cu dopant gradually replaces the Fe atoms starting from the highest potential energy sites. Before the saturation point, the doping process gradually reduces the GB energy (γ), and after the saturation point when Cu starts being accommodated in the bulk region of the grain, the Cu atoms start to replace the Fe atoms outside the grain boundary area, which results in an increase of the GB energy. The equilibrium structure close to the oversaturation point is picked for the later simulations, and this equilibrium structure has the lowest GB energy for each GB configuration discussed in this article.

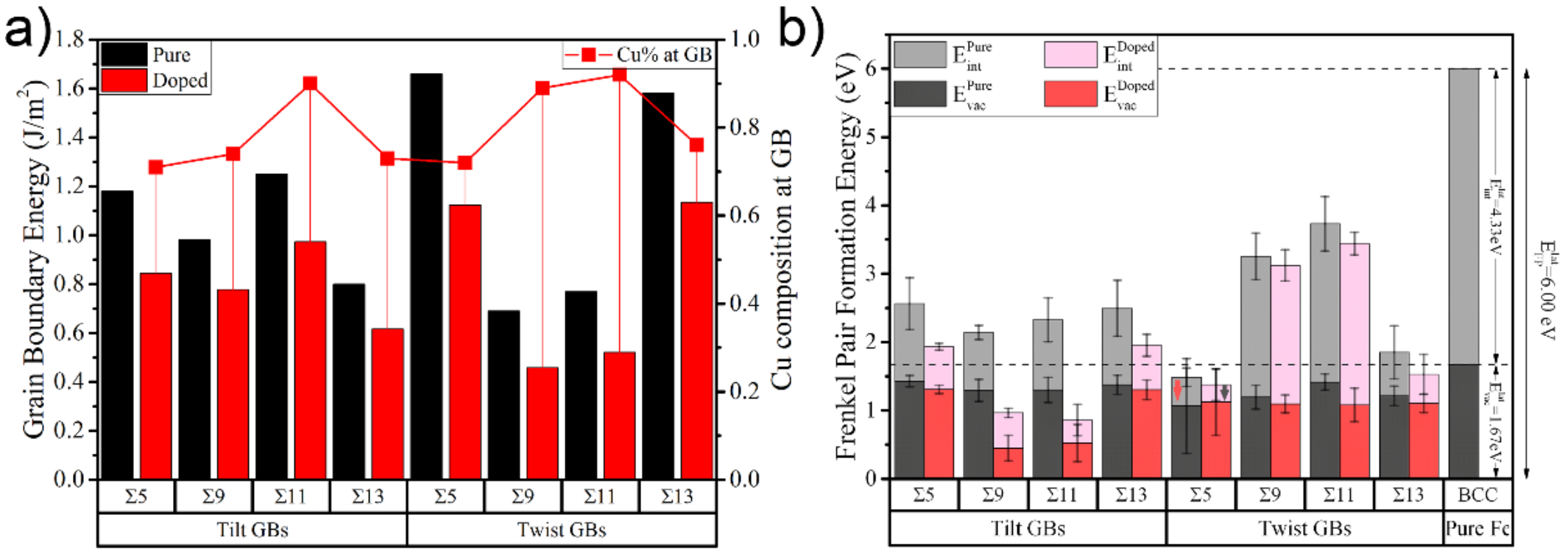

Figure 4: Doping effects on GB energetics. a) Grain boundary energy of each GB structure. The line chart shows the concentration of the Cu in the doped GB. b) The defect formation energy of the interstitial and the vacancy in each GB structure. The last bar shows the defect formation energy in the pure Fe lattice. It is worth noting that the interstitial formation energy at the pure Fe Σ5 twist GB is -0.42 eV, and is -0.25 eV at the doped Σ5 twist GB.

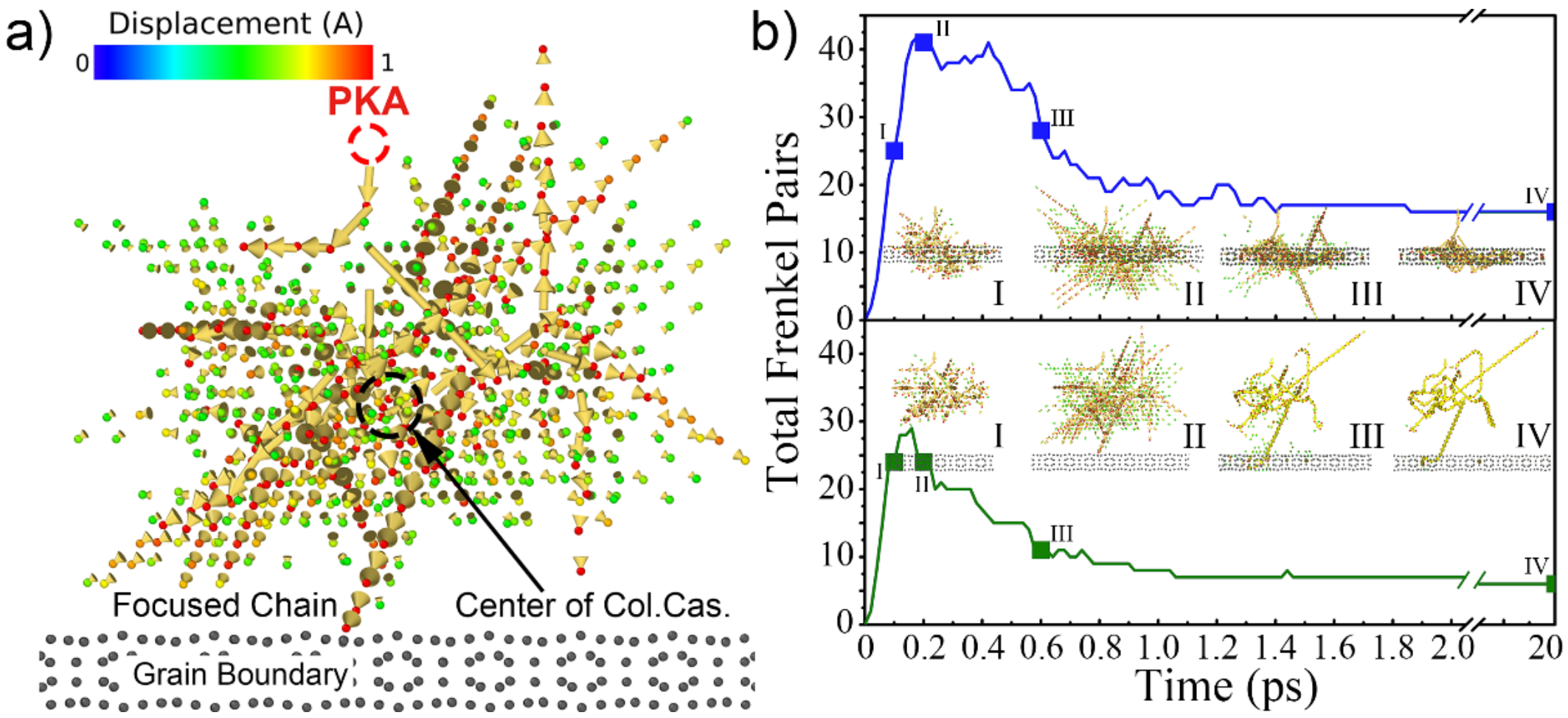

Figure 5: Overlap between collision cascades and grain boundaries. (a) A snapshot of a representative collision cascade around its peak, showing that the center of the collision cascade (CC) does not coincide with the PKA. b) evolution of the collision cascade in two different overlapping scenarios: upper panel, CC overlaps on the grain boundary, lower panel, center of the CC 2nm above the grain boundary. The defect creation processes differ dramatically between these two scenarios. Based on this observation, we decided to use center of the collision cascade as the indicator of the interaction between the Collision Cascade and the GB. The cascade offset is linearly related to the PKA offset.

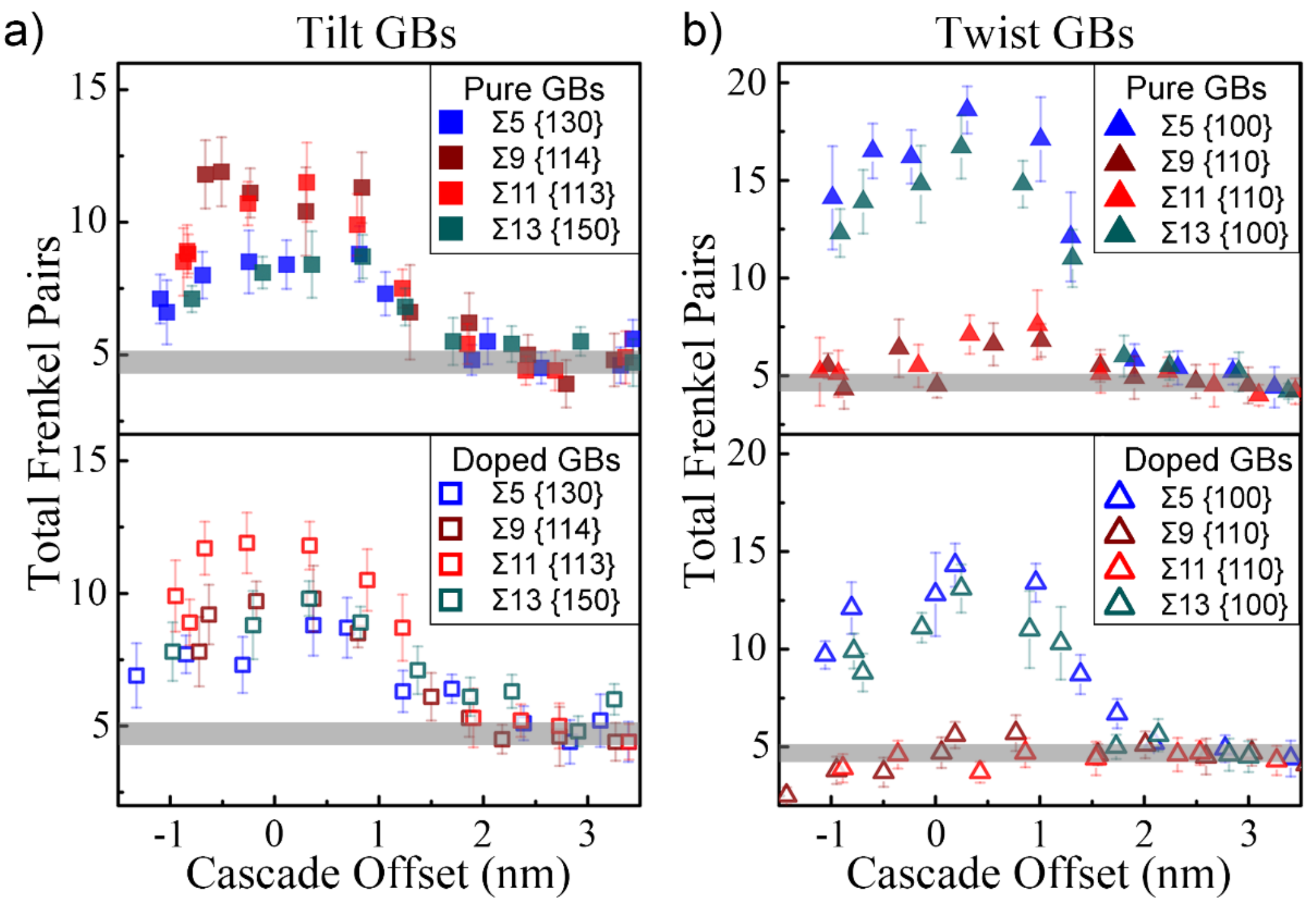

Figure 6: As an effect of the cascade offset, the total number of surviving Frenkel Pairs varied by cascade offset. The number of surviving Frenkel Pairs per CC as a function of the Cascade Offset shows that the defect production rate peaks when the cascade overlaps on the GB. The number of surviving Frenkel Pairs are counted when the CC totally dissipated, which is 20ps after the initiation of the PKA with 1 KeV kinetic energy. When the Cascade Offset is beyond 2nm distance, the trends enter the Bulk region, producing identical amount of Frenkel Pairs as the CC happens in the Bulk structure, which is around 4.7 FPs/CC. Based on these findings, the following discussions will focus on three scenarios: i. Overlap Scenario, where Cascade Offset is 0nm, ii. Intermediate Scenario, where Cascade Offset is 1nm, iii. Indirect Interaction Scenario, where Cascade Offset is 2nm.

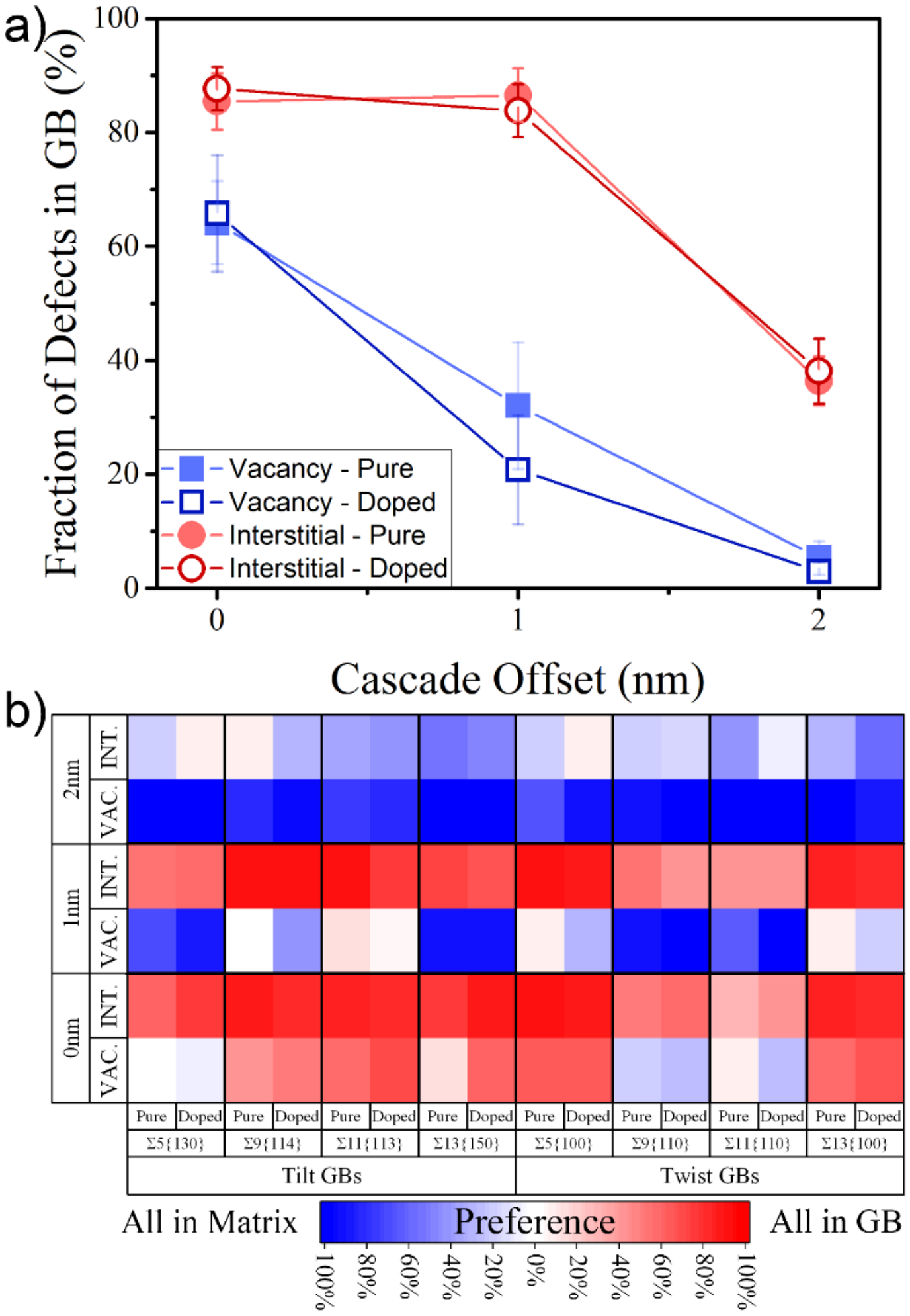

Figure 7: (a) Distribution of the defects showing site preference change at different cascade offsets. With different CC-GB interaction, the defect distributed differently. To take a closer look, (b) a defect distribution map has been generated, showing biased effect of the grain boundary during the defect creation process at different collision cascade offsets. At 0nm offset, which implies the most intensive CC-GB interaction: the interstitials had a generally high preference to form in the grain boundary area; for four out of eight grain boundaries (tilt Σ9, tilt Σ11, Σ5 twist and twist Σ13), more residual vacancies were left in the grain boundary area, but for samples with the tilt Σ5, Σ9 twist or Σ11 twist GBs, the distribution of the vacancies didn't have the region preference as strong as other samples. At 1nm cascade offset, the separation of the interstitial and vacancy distribution started to show up, the grain boundary retained most of the interstitials compared to the case of 0nm offset, but significantly higher percentage of the vacancies were left in the grain lattice. When the cascade core moved further away from the grain boundary to 2nm offset, most of the vacancies were left in the grain lattice region in all samples but the interstitials still had a relatively high probability to be found in the grain boundary area.

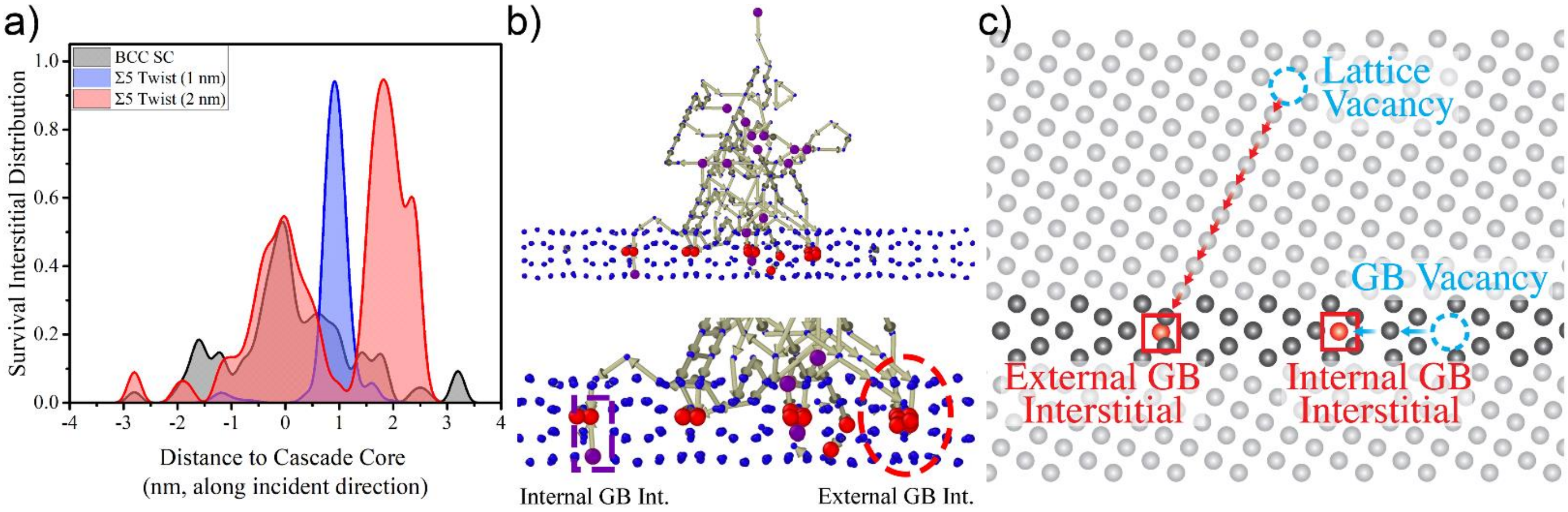

Figure 8: (a) The interstitial distribution as a function of distance to cascade core in the BCC pure Fe single crystal and in the pure Fe structure with a Σ5 twist GB. The SC structure shares the same crystallographic orientation with the upper grain of the Σ5 twist structure where the PKA initiated in. In the Σ5 twist structure, the cascade core is 1 or 2 nm away from the GB plane. This shows how the survival interstitials distribute along the incident direction of the PKA. The interstitials produced by the cascade near Σ5 twist GB are mostly in the GB, showing a narrow distribution, while the interstitial distribution in the single crystal is much broader and is symmetrical along centerplate. (b) the final atomic displacement state after the dissipation of the cascade, the big red spheres marked all the interstitials left, and the small purple spheres marked all the vacancies. Depending on the origin of the interstitial atom, the grain boundary interstitials were classified into the external GB interstitials and the internal GB interstitials. (c) schematics of two different types of grain boundary interstitials. In the external GB interstitial configuration, a lattice vacancy is linked to the grain boundary interstitial by the replacement sequencies. The mass transfer from the grain lattice to the grain boundary happens during the process of the external GB interstitial creation. In the internal GB interstitial configuration, the interstitial atom comes from the nearby grain boundary site, the grain boundary Frenkel pair is created by the displacement of the grain boundary atom.

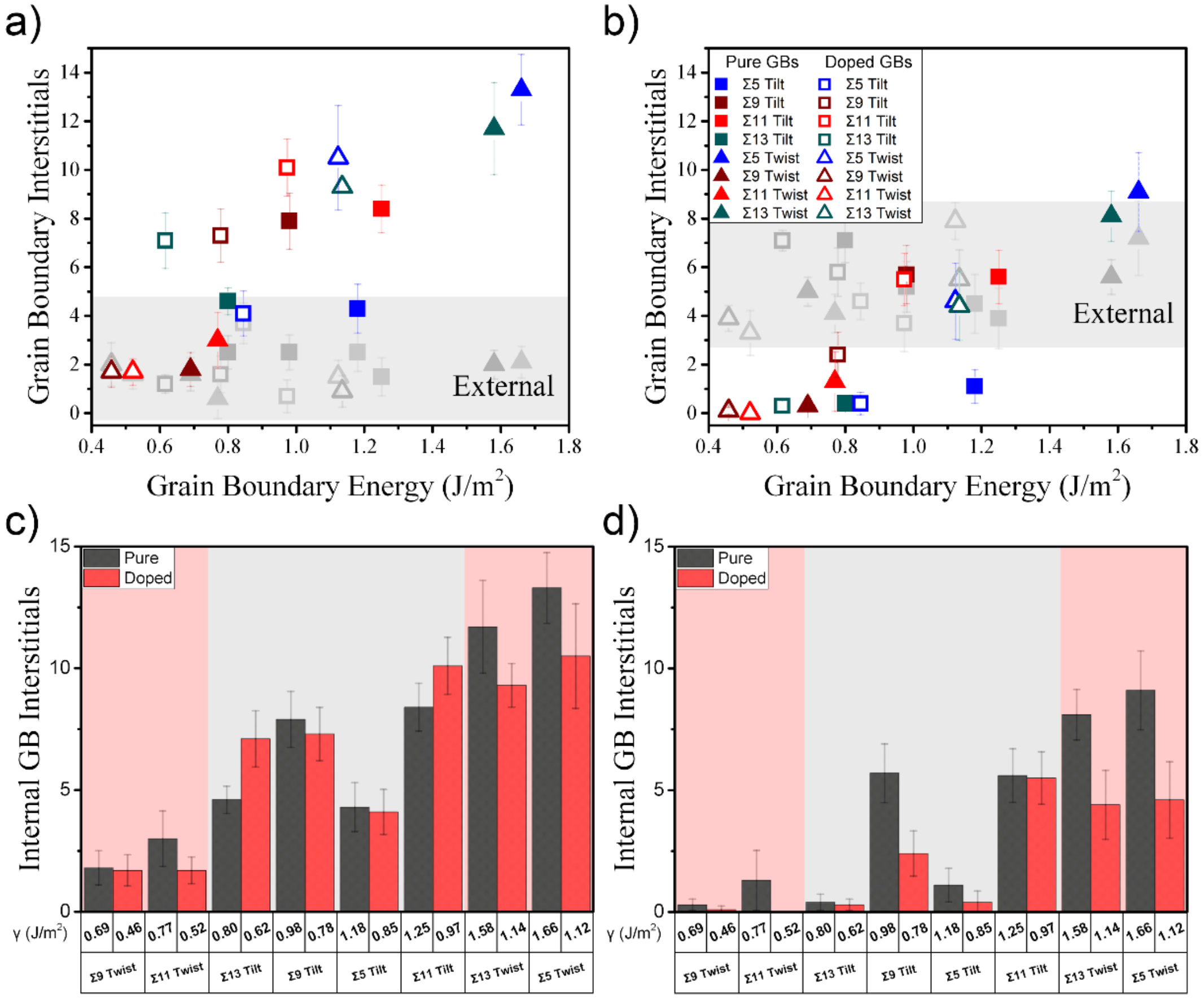

Figure 9: The effect of grain boundary energy on grain boundary interstitial production at different offsets. a) The GB interstitial production as a function of grain boundary energy at 0 nm cascade offset. b) The GB interstitial production as a function of grain boundary energy at 1 nm cascade offset. c) The doping effect on the internal GB interstitial production rate at 0 nm cascade offset. d) The doping effect on the internal GB interstitial production rate at 1 nm cascade offset.

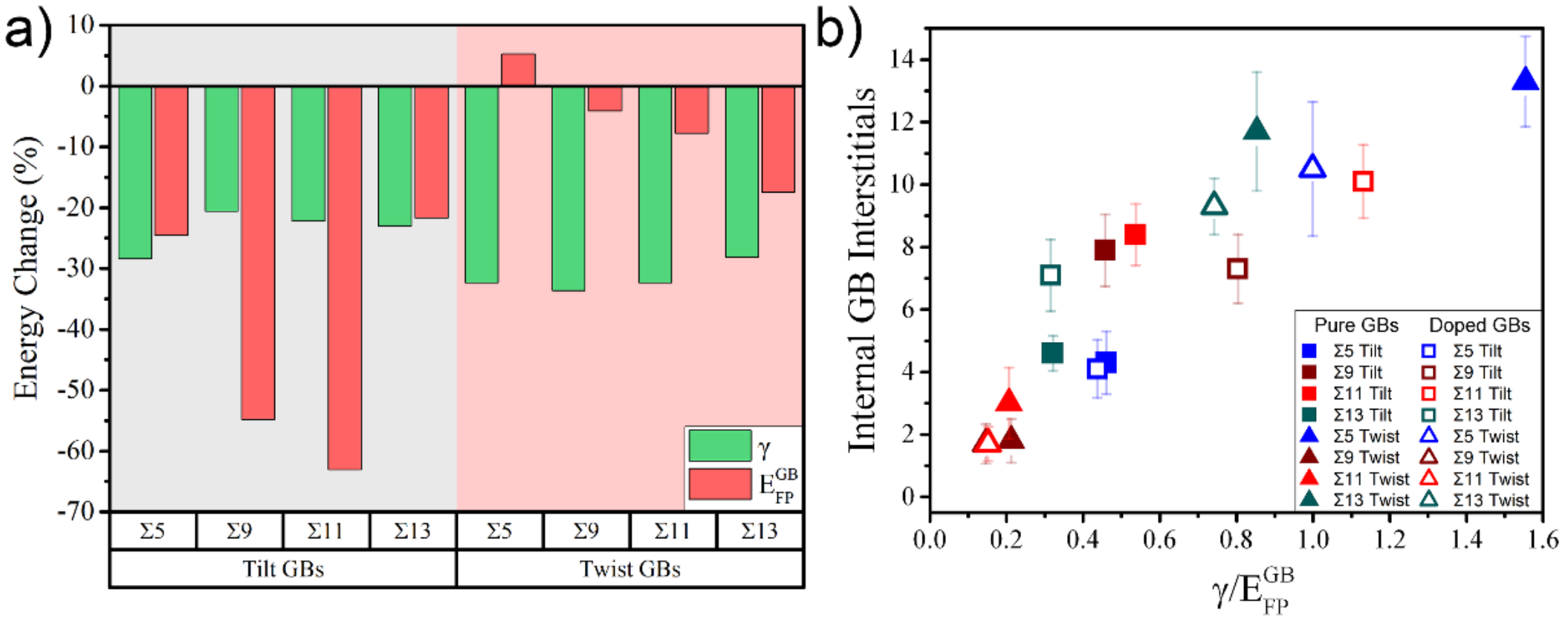

Figure 10: Probable multi-factor dependence of the defect production process (grain boundary energy and grain boundary Frenkel pair formation energy). a) Percentage change of the grain boundary energy (green bars) and the GB Frenkel pair formation energy after doping in the tested grain boundaries. Doping can have different influences on these two key parameters related to the defect production. b) the dependence of the internal GB interstitial production on the combined parameter ($^{\gamma}/_{E_{FP}^{GB}}$) was showed. The internal GB interstitials to the combined parameter shows a higher correlation coefficient (0.895) than that of the internal GB interstitials solely to either grain boundary energy (0.817) or the GB Frenkel pair formation energy (-0.822).

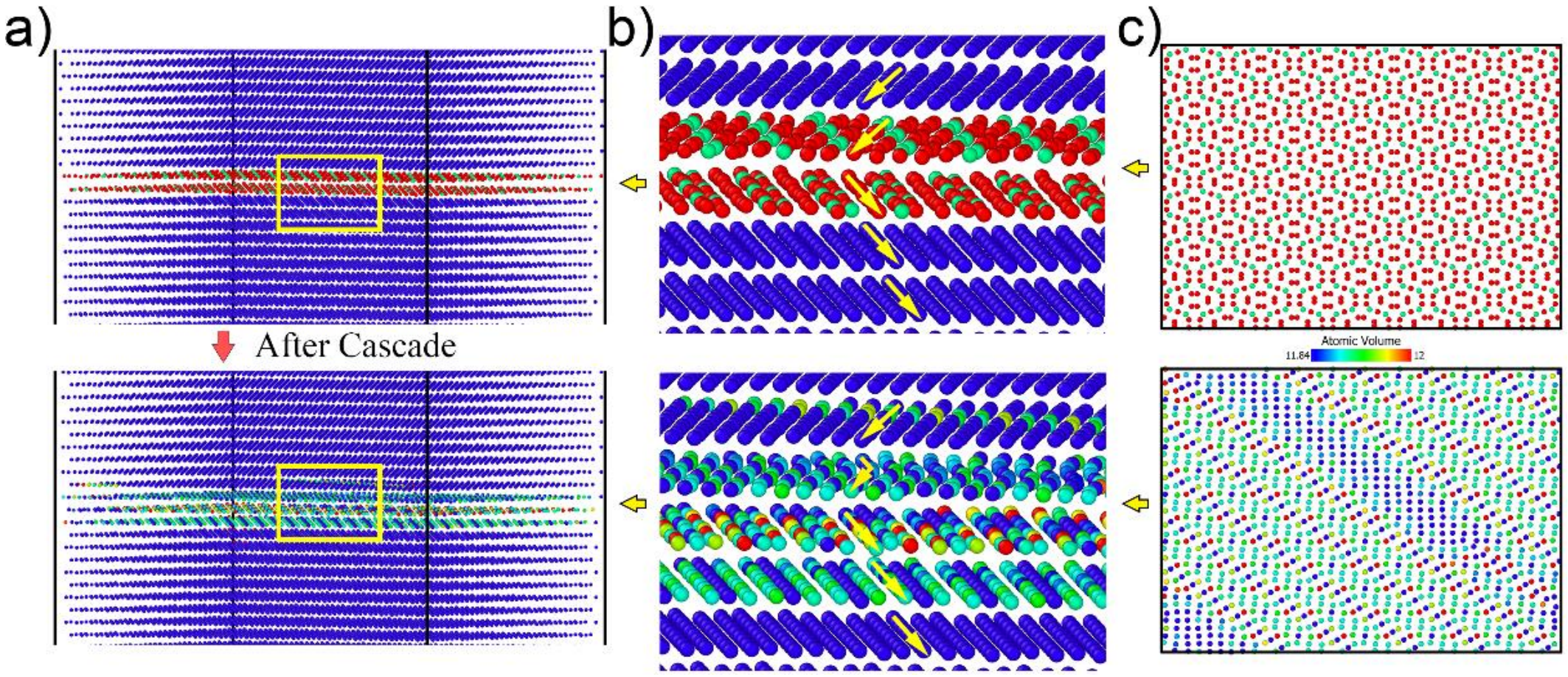

Figure 11: The cascade on the Twist Σ9 pure Fe GB can induce grain boundary reconfiguration, and this type of reconfiguration reduces the total energy of the Twist Σ9 GB. a) the Twist Σ9 grain boundary structure before and after the cascade. b) the zoom-in view of the structure. c) the top view on the grain boundary plane.

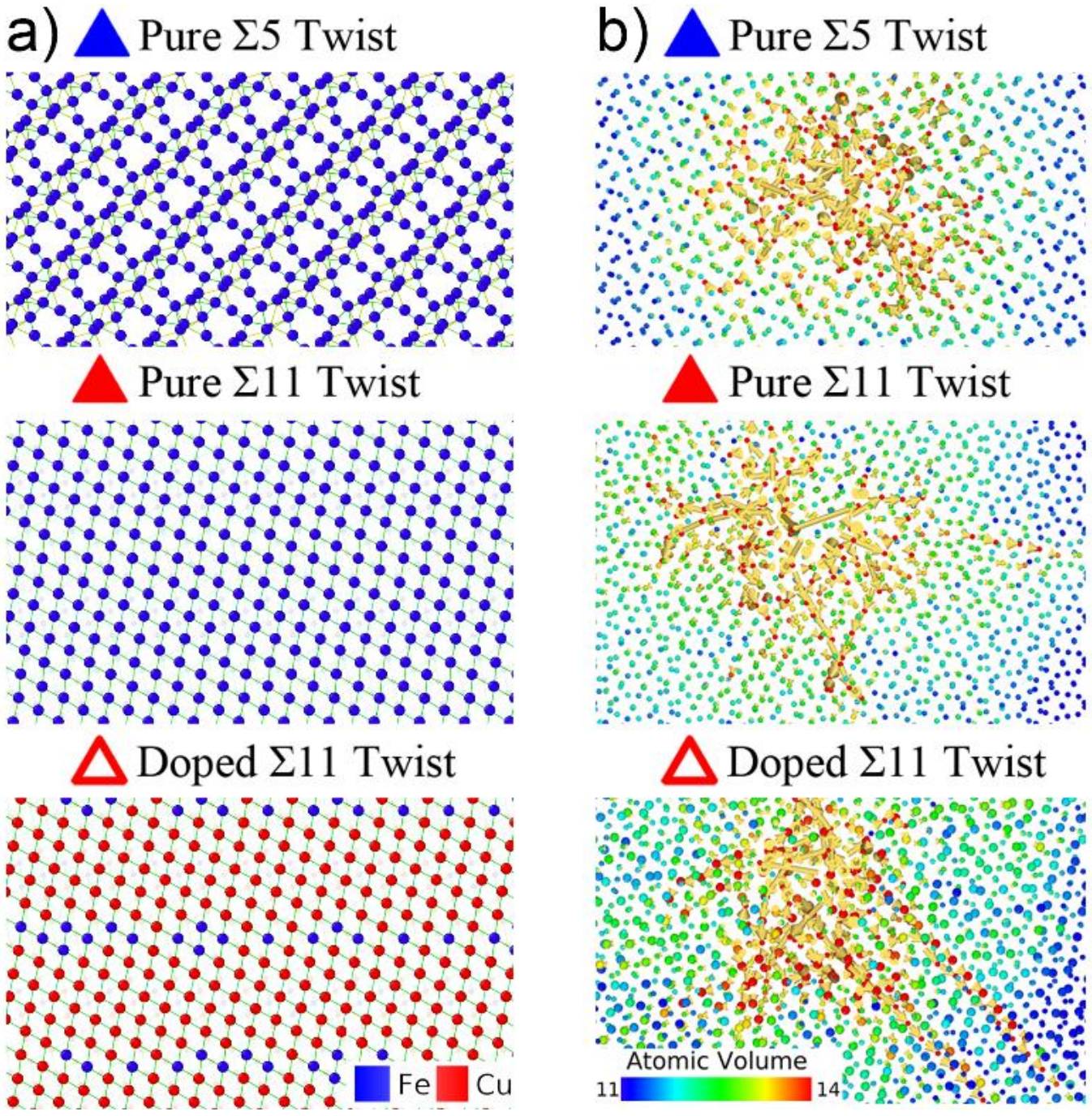

Figure 12: GB structure effect on cascade evolution at 0 nm cascade offset. a) the pristine grain boundary configurations under the perpendicular perspective to the grain boundary plane. b) the collision cascade at its maximum size at different grain boundaries. At the grain boundary with a rough GB plane configuration (Σ5 twist GB), the displacements cannot initiate long sequences. The formation of focused chains during the cascade evolution was suppressed. At the grain boundary with a smooth plane (Twist Σ11 GB), the displacements can travel a relatively long distance forming the displacement sequences through the focusing mechanism. In the present paper, those displacement sequences formed during the evolution were noted as the focused chains to distinguish from the replacement collision sequences which emphasize the final state of the displacements.

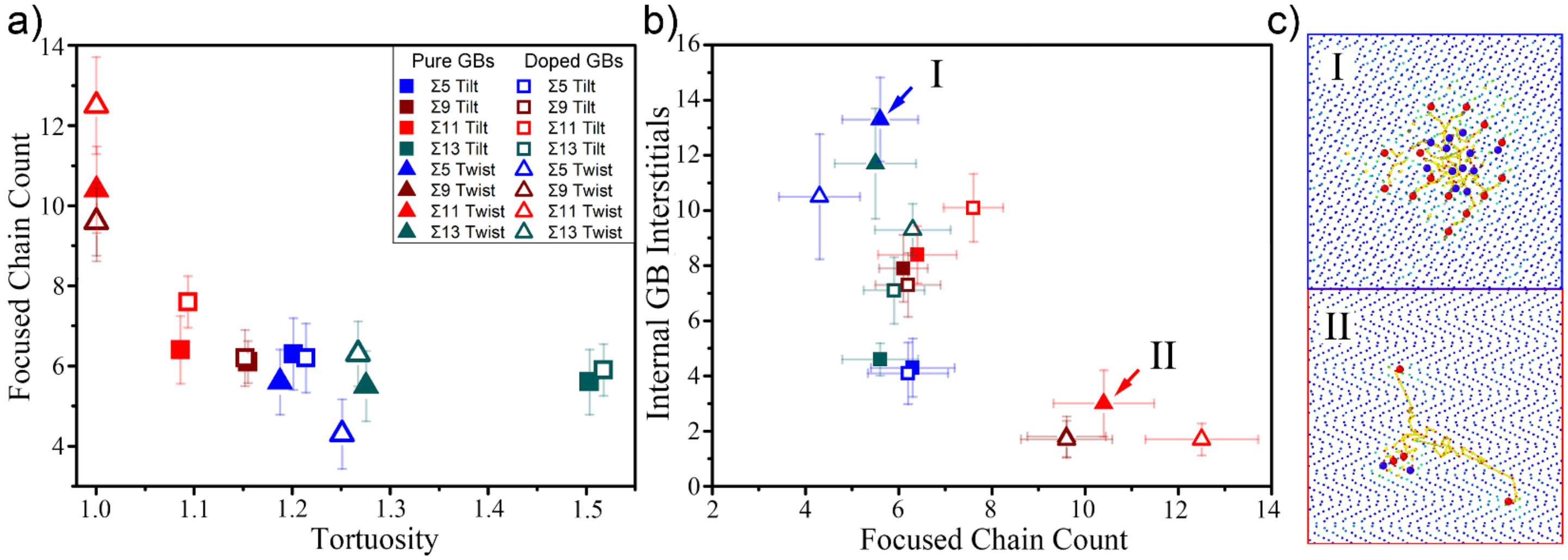

Figure 13: Probable structure dependent mechanism related to the reduction of GB defect production in the unique GBs. a) the number of the focused chain events during the cascade as a function of the GB plane tortuosity. b) the production of the GB interstitials as a function of the number of the focused chain events. c) I. shows the final snapshot of the cascade developed at the pure Fe Σ5 twist GB. II. shows the final snapshot at the Twist Σ11 pure Fe GB.

**Tables**

| GB Type | $\gamma_{pure}$ (J/m$^2$) | $\gamma_{doped}$ (J/m$^2$) | $X_{Cu}^{gb}$ |
|---|---|---|---|
| **Σ5 Tilt** | 1.18 | 0.85 | 0.71 |
| **Σ9 Tilt** | 0.98 | 0.78 | 0.74 |
| **Σ11 Tilt** | 1.25 | 0.97 | 0.90 |
| **Σ13 Tilt** | 0.8 | 0.62 | 0.73 |
| **Σ5 Twist** | 1.66 | 1.12 | 0.72 |
| **Σ9 Twist** | 0.69 | 0.46 | 0.89 |
| **Σ11 Twist** | 0.77 | 0.52 | 0.92 |
| **Σ13 Twist** | 1.58 | 1.14 | 0.76 |

Table 1: Grain boundary energy of the pure Fe GBs and the doped GBs, and the Cu concentration in the grain boundary region for each doped grain boundary.

**Supplementary**

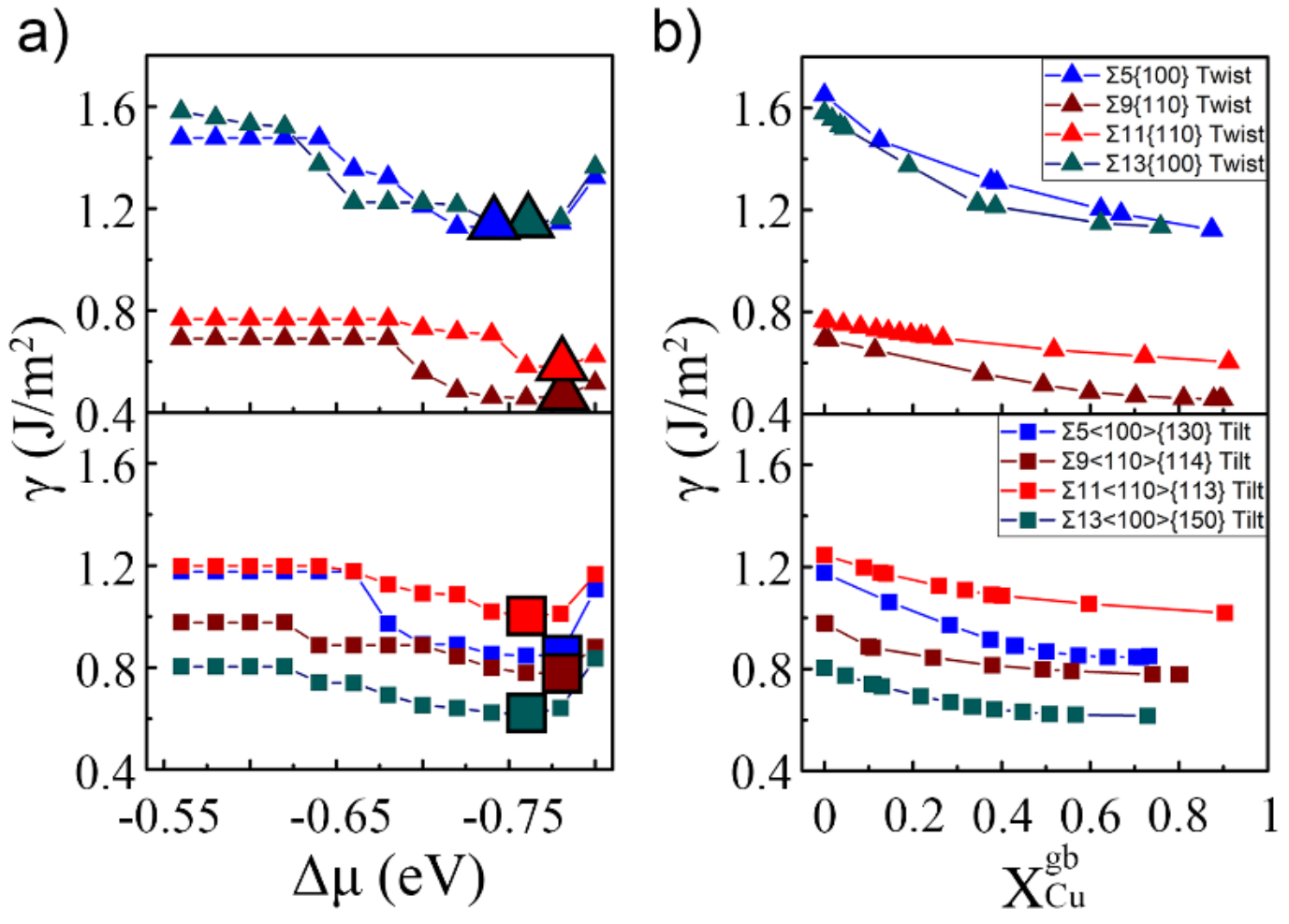

Sup. 1: a) grain boundary energy as a function of Δμ. b) grain boundary energy as a function of the Cu concentration in the grain boundary region.

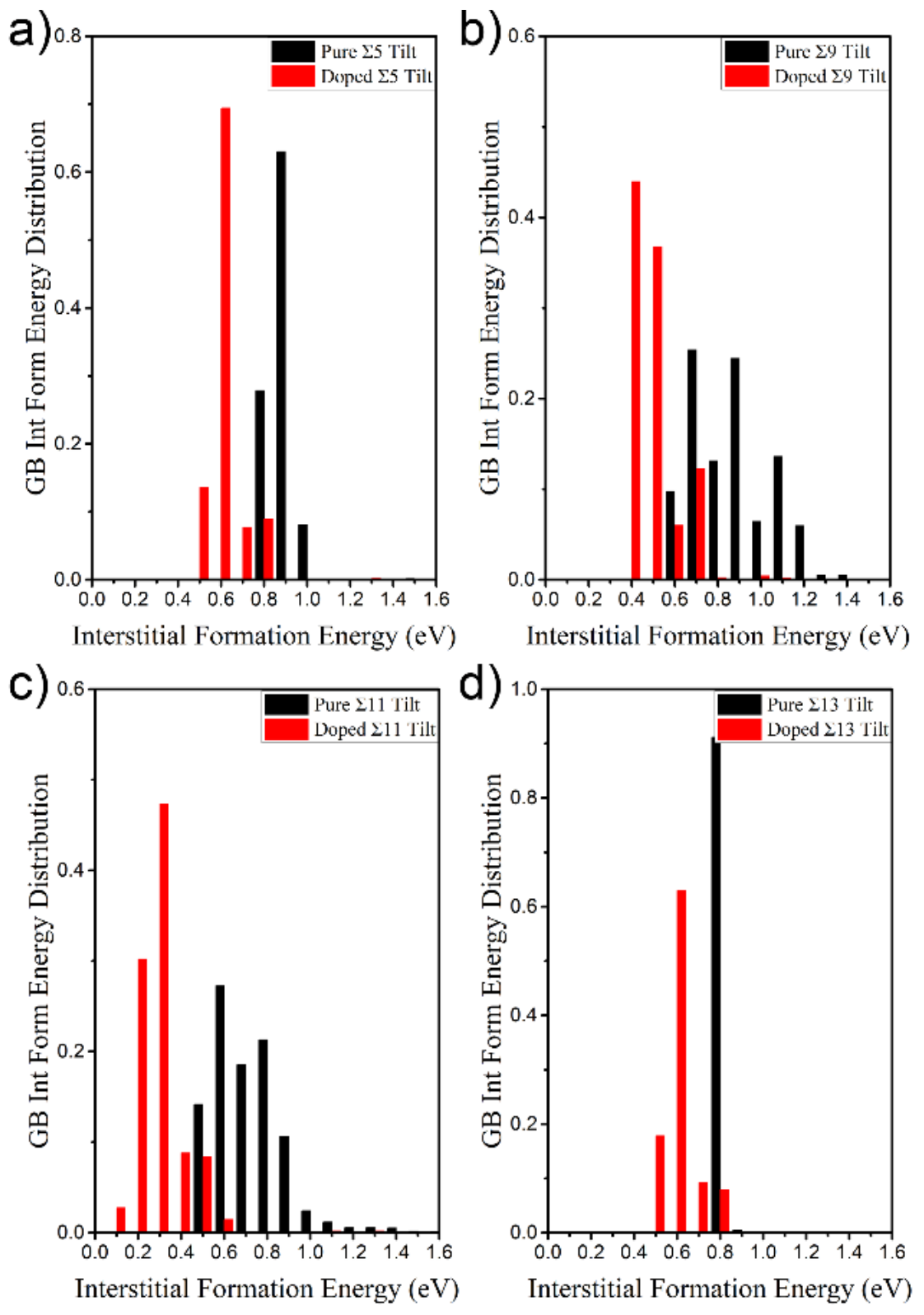

Sup. 2: Interstitial formation energy distribution at tilt GBs: a) Tilt Σ5 GB, b) Tilt Σ9 GB, c) Tilt Σ11 GB, and d) Tilt Σ13 GB.

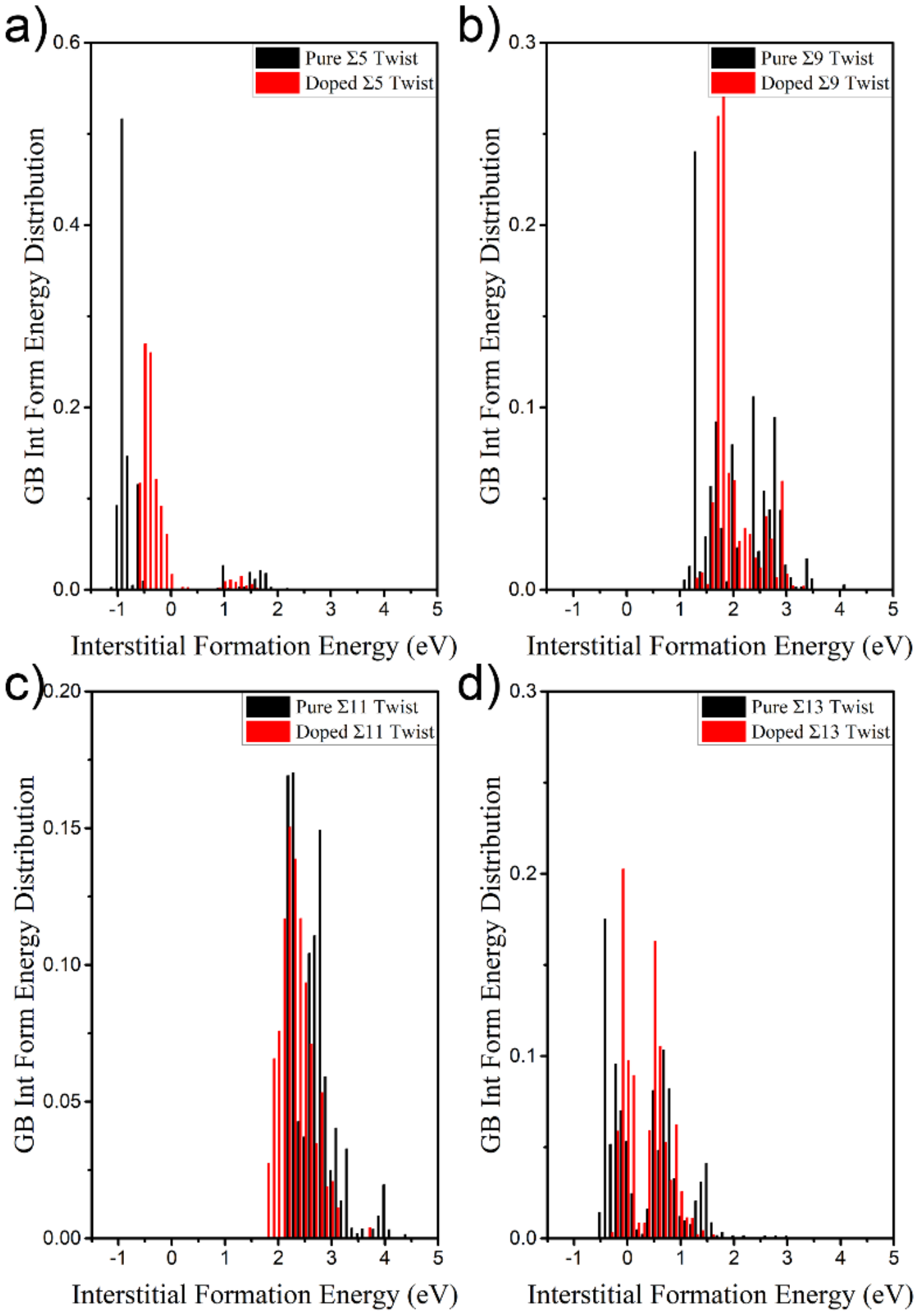

Sup. 3: Interstitial formation energy distribution at twist GBs: a) Twist Σ5 GB, b) Twist Σ9 GB, c) Twist Σ11 GB, and d) Twist Σ13 GB.

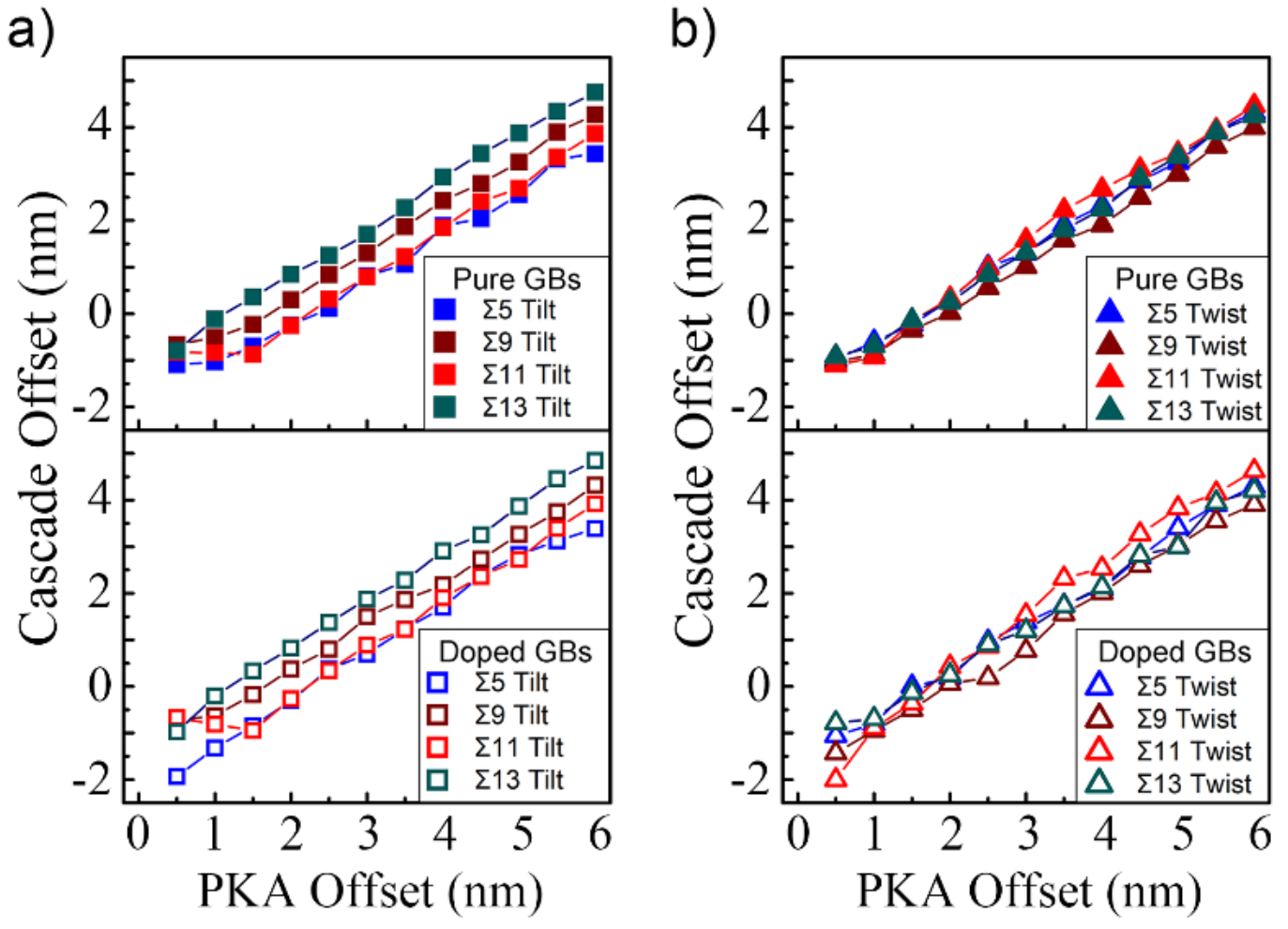

Sup. 4: The correlation between the PKA offset and the cascade offset in all tested structures.